\documentclass[%
 reprint,
 amsmath,amssymb,
 aps,
pra,
]{revtex4-2}

\usepackage{graphicx}
\usepackage{dcolumn}
\usepackage{bm}
\usepackage{physics}
\usepackage{xcolor}
\usepackage{makecell}

\begin{document}

\preprint{APS/123-QED}

\title{Quantum dynamics of molecular ensembles coupled with quantum light: Counter-rotating interactions as an essential component} 

\author{Yi-Ting Chuang}
\affiliation{%
Department of Chemistry, National Taiwan University, Taipei 10617, Taiwan
}%
\affiliation{Institute of Atomic and Molecular Sciences, Academia Sinica, Taipei 10617, Taiwan}

\author{Liang-Yan Hsu}%
\email{lyhsu@gate.sinica.edu.tw}
\affiliation{%
Institute of Atomic and Molecular Sciences, Academia Sinica, Taipei 10617, Taiwan
}
\affiliation{%
Department of Chemistry, National Taiwan University, Taipei 10617, Taiwan
}
\affiliation{National Center for Theoretical Sciences, Taipei 10617, Taiwan}%

\date{\today}

\begin{abstract}
The rotating-wave approximation to light-matter interactions is widely used in the quantum electrodynamics Hamiltonian; however, its validity has long been a matter of debate. In this article, we explore the impact of the rotating-wave approximation on the quantum dynamics of multiple molecules in complex dielectric environments within the framework of macroscopic quantum electrodynamics. In general, we find that the energy shifts of the molecules and the inter-molecule dipole-dipole interaction obtained in the weak coupling regime are correct only when the counter-rotating interactions are considered. Moreover, under the rotating-wave approximation, the energy shifts of the ground-state molecules and a portion of the inter-molecule interaction are discarded. Notably, in the near-field zone (short inter-molecular distance), the reduction of inter-molecule interaction can reach up to 50 percent. We also conduct a case study on the population dynamics of a pair of identical molecules above a plasmonic surface. Through analytical and numerical analysis, it is revealed that the rotating-wave approximation can profoundly affect the dynamics of the molecules in both strong and weak coupling regimes, emphasizing the need for careful consideration when making the rotating-wave approximation in a multiple-molecule system coupled with quantum light.    
\end{abstract}

\maketitle

\section{Introduction}
Over the past few decades, extensive experimental and theoretical research has demonstrated that coupling molecules (or quantum emitters) with confined electromagnetic fields in a well-designed photonic environment is a promising avenue for modifying physical processes such as spontaneous emission \cite{Bonifacio1971,Bonifacio1971b,Morawitz1979,Seke1986,Seke1986b,Kulina1986,Bashkirov1990,Alber1992,Kozierowski1995,Temnov2005,Martín2010,Goban2015,Zhang2016,Kim2018, Luo2019,Wang2023} and resonance energy transfer \cite{Joel1984,Joel1985,Hua1985,Kobayashi1995,Andrew2000,Zhang2007,Govorov2007,Reil2008,Lessard2009,Lunz2011,Zhong2017,Ren2017,Wu2018}. Recently, there has been a surge of interest in utilizing the concept of coupling molecules with confined electromagnetic fields to manipulate chemical reactions \cite{Hutchison2012,Thomas2016,Thomas2019,Lather2019,Vergauwe2019,Thomas2020,Sau2021}. The successful realization of such experiments has spurred scientists to develop theoretical frameworks aimed at elucidating the underlying physics behind observed phenomena and predicting novel effects. Among these theoretical frameworks, cavity quantum electrodynamics (QED) Hamiltonian are extensively used, especially the Tavis-Cummings model and its extensions \cite{Galego2015,Herrera2016,Wu2016,Ribeiro2018,Semenov2019,Spano2020,Du2022}.

Qunatum electrodynamics serves as one of the most favorable foundations to study the interactions between molecules and electromagnetic fields. Within the framework of QED, one typically needs to make use of further approximations to simplify the light-matter interactions, and the rotating-wave approximation (RWA) \cite{Scully1997,Vogel2006} is viewed as one of the most commonly adopted approximations. To simply demonstrate the key concept of the RWA, we consider a simple model consisting of a single molecule ($\omega_\mathrm{m}$) interacting with a single photonic mode ($\omega_\mathrm{p}$), i.e.,
\begin{align}
\nonumber
    \hbar\omega_\mathrm{m} \hat{\sigma}^{(+)}  \hat{\sigma}^{(-)} + \hbar\omega_\mathrm{p} \hat{a}^\dagger  \hat{a} + \hbar g \left[ \hat{\sigma}^{(+)} +  \hat{\sigma}^{(-)} \right] \left[ \hat{a}^\dagger +  \hat{a} \right].
\end{align}
The spirit of the RWA lies in retaining the co-rotating interaction $\hbar g \left[ \hat{\sigma}^{(+)} \hat{a} + \hat{\sigma}^{(-)} \hat{a}^\dagger \right]$ while disregarding the counter-rotating interaction $\hbar g \left[ \hat{\sigma}^{(+)} \hat{a}^\dagger + \hat{\sigma}^{(-)} \hat{a} \right]$. This choice is motivated by the fact that the co-rotating interaction oscillates at a relatively low frequency $\omega_\mathrm{m} - \omega_\mathrm{p}$ while the counter-rotating interaction oscillates at a higher frequency  $\omega_\mathrm{m} + \omega_\mathrm{p}$. By neglecting the counter-rotating interaction, the RWA simplifies the mathematical treatment of the system and allows for an easier analysis. Typically, the RWA is considered a good approximation when (i) the light-matter coupling strength $g$ is weak and (ii) the detuning $\omega_\mathrm{m} - \omega_\mathrm{p}$ is small. However, when molecules are coupled to multiple photonic modes (including off-resonant modes), such as in free space (infinite photonic modes), condition (ii) seems to be violated anyhow.

The validity of the RWA has long been a matter of debate. While there has been extensive research on the impact of the RWA, it is worth noting that most of these studies have primarily focused on scenarios involving only a single photonic mode \cite{Zaheer1988, Klimov2001,Larson2007,Irish2007, Werlang2008,Zueco2009}, in free space \cite{Agarwal1971,Knight1973,Friedberg2008,Fleming2010}, or in non-dispersive and non-absorbing (lossless) media \cite{Jørgensen2022}. Nevertheless, in experimental setups, such as the Fabry-Pérot cavity, there exist infinite numbers of photonic modes, including many off-resonant modes \cite{Wei2021}, that are dressed by the surrounding photonic environments. The interactions between molecules and these dressed photons (polaritons) are supposed to play a crucial role in determining the chemical and physical properties of the system. Therefore, restricting our analysis to a single photonic mode or the free space scenario may result in misinterpretations of experimental observations and impede our ability to accurately predict and quantify polariton-coupled processes. In addition, it has been shown that plasmonic materials can strongly affect physical processes, including spontaneous emission \cite{Wang2020, Wang2023} and resonance energy transfer \cite{Hsu2017,Wu2018}, and it is natural to consider dispersive and absorbing media in studying plasmonic effects. To overcome the aforementioned issues, one needs advanced theoretical treatments that incorporate the coupling of molecules with infinite photonic modes and take into account the effects of (dispersive and absorbing) photonic environments.

In this study, in order to demonstrate counter-rotating interactions as an essential component, we investigate the quantum dynamics of multiple molecules in dielectric environments using macroscopic quantum electrodynamics (MQED) \cite{Gruner1996,Dung1998,Scheel1998}, which is an effective field theory for describing quantized electromagnetic fields in any arbitrary inhomogeneous, dispersive, and absorbing dielectric environment. We elucidate the importance of counter-rotating interactions in the quantum dynamics of
multiple molecules in both strong and weak coupling regimes. In addition, we clarify that a reduction of the resonant dipole-dipole interaction due to the RWA has been wrongly interpreted as a non-negligible (quantum) effect in previous works \cite{Dzsotjan2011,Varguet2021}. Note that our theoretical approach is not only restricted to the study of the RWA, but can also be applied to investigate the combined effect of molecular fluorescence and excitation energy transfer in any arbitrary photonic environments, holding great promise for further advancements in areas such as quantum optics, nanophotonics, and molecular engineering.

This article is organized as follows. In Sec.~\ref{Sec:Theory}, starting from the MQED Hamiltonian (including without and with the use of the RWA), we derive the dynamical equations that can describe the quantum dynamics of multiple emitters in complex dielectric environments. In addition, the underlying physics behind the dynamical equations is discussed. In Sec.~\ref{Sec:WeakCoupling}, we apply the Markov approximation to the dynamical equations to study their behavior in the weak coupling regime. In this regime, we obtain the energy shift, decay rate, and inter-molecule dipole-dipole interaction, and compare our results to the previous works. In Sec.~\ref{Sec:NumericalStudy}, we apply our approach to investigate the role of counter-rotating interactions in the quantum dynamics of a pair of identical molecules above a plasmonic surface in both strong and weak coupling regimes. In the last section, we provide a concise summary of this study. 

\section{Theory}
\label{Sec:Theory}
\subsection{Hamiltonian}
Considering a collection of two-level molecules coupled to polaritons (dressed photons) in an arbitrary inhomogeneous, dispersive, and absorbing medium, the total Hamiltonian $\hat{H}$ (without the RWA) under the electric-dipole approximation in the multipolar coupling MQED \cite{Buhmann2012,Vogel2006} can be expressed in terms of the molecular Hamiltonian $\hat{H}_\mathrm{M}$, polaritonic Hamiltonian $\hat{H}_\mathrm{P}$ and interaction Hamiltonian $\hat{H}_\mathrm{I}$ as
\begin{align}
    \hat{H} = \hat{H}_\mathrm{M} + \hat{H}_\mathrm{P} + \hat{H}_\mathrm{I},
\label{Eq:Hamiltonian_Tot}
\end{align}
with
\begin{align}
    \hat{H}_\mathrm{M} &= \sum_{\alpha} \hbar\omega_{\alpha} \hat{\sigma}^{(+)}_{\alpha} \hat{\sigma}^{(-)}_{\alpha},
    \label{Eq:Hamiltonian_Mol} \\
    \hat{H}_\mathrm{P} &= \int \mathrm{d}\mathbf{r} \int_{0}^{\infty} \mathrm{d}\omega \, \hbar\omega \,\mathbf{\hat{f}}^\dagger(\bf{r},\omega)\cdot\mathbf{\hat{f}}(\bf{r},\omega),
    \label{Eq:Hamiltonian_Pol} \\
     \hat{H}_\mathrm{I} &= -\sum_{\alpha} \hat{\boldsymbol{\mu}}_\alpha \cdot \hat{\mathbf{F}}(\mathbf{r}_\alpha).
    \label{Eq:Hamiltonian_Int}
\end{align}
The molecular Hamiltonian $\hat{H}_\mathrm{M}$ in Eq.~(\ref{Eq:Hamiltonian_Mol}) describes the total energy of the molecules, where $\omega_\mathrm{\alpha}$, $\hat{\sigma}^{(+)}_{\alpha}$ and $\hat{\sigma}^{(-)}_{\alpha}$ denote the electronic transition frequency, raising operator and lowering operator of the $\alpha$-th molecule, respectively. The raising and lowering operators of the $\alpha$-th molecule can be defined using the electronically ground state $\ket{\mathrm{g}_\alpha}$ and excited state $\ket{\mathrm{e}_\alpha}$ of $\alpha$ as follows: $\hat{\sigma}^{(+)}_{\alpha} = \ket{\mathrm{e}_\alpha}\bra{\mathrm{g}_\alpha}$ and $\hat{\sigma}^{(-)}_{\alpha}$ = $\ket{\mathrm{g}_\alpha} \bra{\mathrm{e}_\alpha}$. Note that we have neglected the dipole self-interaction since it only contributes to a small energy shift (free space Lamb shift) after renormalization \cite{Power1959}. The polaritonic Hamiltonian $\hat{H}_\mathrm{P}$ in Eq.~(\ref{Eq:Hamiltonian_Pol}) describes the energy of polaritons in the dielectric environment, where $\mathbf{\hat{f}}^\dagger(\mathbf{r},\omega)$ and $\mathbf{\hat{f}}(\mathbf{r},\omega)$ are the creation and annihilation operators of the bosonic vector fields that satisfy the commutation relations:
\begin{align}
    \nonumber                      \left[ \hat{f}_k\left( \mathbf{r}, \omega \right), \hat{f}^\dagger_{k'}\left( \mathbf{r}', \omega' \right) \right] &= \delta_{kk'} \delta\left( \mathbf{r}   - \mathbf{r}' \right) \delta\left( \omega - \omega' \right), \\
    \nonumber                      \left[ \hat{f}_k\left( \mathbf{r}, \omega \right), \hat{f}_{k'}\left( \mathbf{r}', \omega' \right) \right] &= 0.
\end{align}
The interaction Hamiltonian $\hat{H}_\mathrm{I}$ in Eq.~(\ref{Eq:Hamiltonian_Int}) describes the couplings between the molecules and polaritons including counter-rotating interactions, i.e., no RWA , where $\hat{\boldsymbol{\mu}}_\alpha$ and $\hat{\mathbf{F}}(\mathbf{r}_\alpha)$ are the transition dipole operator of $\alpha$ and the field operator, respectively. The transition dipole operator $\hat{\boldsymbol{\mu}}_\alpha$ can be written in terms of $\hat{\sigma}^{(+)}_{\alpha}$ and $\hat{\sigma}^{(-)}_{\alpha}$ as follows, 
\begin{align}
    \hat{\boldsymbol{\mu}}_\alpha = \boldsymbol{\mu}^\mathrm{eg}_\alpha \hat{\sigma}^{(+)}_{\alpha} + \boldsymbol{\mu}^\mathrm{ge}_\alpha \hat{\sigma}^{(-)}_{\alpha},
\end{align}
where $\boldsymbol{\mu}^\mathrm{eg}_\alpha = (\boldsymbol{\mu}^\mathrm{ge}_\alpha)^*$ is the electronic transition dipole moment of $\alpha$. The field operator is defined as:
\begin{align}
    \hat{\mathbf{F}}(\mathbf{r}_\alpha) =  \hat{\mathbf{F}}^{(+)}(\mathbf{r}_\alpha) + \mathrm{H.c.},
\end{align}
with
\begin{gather}
    \hat{\mathbf{F}}^{(+)}(\mathbf{r}_\alpha) = \int_0^\infty \dd{\omega}   \int \dd{\mathbf{r}} \overline{\overline{\mathbf{g}}}(\mathbf{r}_\alpha,\mathbf{r},\omega) \cdot \hat{\mathbf{f}}(\mathbf{r},\omega), \\
    \overline{\overline{\mathbf{g}}}(\mathbf{r}_\alpha,\mathbf{r},\omega) = i\sqrt{\frac{\hbar}{\pi\varepsilon_0}} \frac{\omega^2}{c^2} \sqrt{\mathrm{Im} \left[ \varepsilon_\mathrm{r}(\mathbf{r},\omega) \right]} \, \overline{\overline{\mathbf{G}}}(\mathbf{r}_\alpha,\mathbf{r},\omega),
\end{gather}
where $\varepsilon_0$, $\varepsilon_\mathrm{r}(\mathbf{r},\omega)$, and $c$ are the permittivity of free space, relative permittivity, and speed of light in vacuum, respectively.  $\overline{\overline{\mathbf{g}}}(\mathbf{r}_\alpha,\mathbf{r},\omega)$ is an auxiliary tensor defined in terms of the dyadic Green's function $\overline{\overline{\mathbf{G}}}(\mathbf{r}_\alpha,\mathbf{r},\omega)$ that satisfies macroscopic Maxwell’s equations, i.e.,
\begin{equation}
\nonumber
    \left[ \frac{\omega^2}{c^2}\varepsilon_\mathrm{r}(\mathbf{r}_\alpha,\omega) - \nabla \times \nabla \times \right] \overline{\overline{\mathbf{G}}}(\mathbf{r}_\alpha,\mathbf{r},\omega) = -\mathbf{\overline{\overline{I}}}_3 \delta(\mathbf{r}_\alpha-\mathbf{r}),
\label{Eq:GreensFunction}
\end{equation}
where $\mathbf{\overline{\overline{I}}}_3$ and $\delta(\mathbf{r}_\alpha-\mathbf{r})$ are the $3\times3$ identity matrix and three-dimensional delta function, respectively. Note that the dyadic Green's function can be further decomposed as $\overline{\overline{\mathbf{G}}}(\mathbf{r},\mathbf{r}',\omega) = \overline{\overline{\mathbf{G}}}_0(\mathbf{r},\mathbf{r}',\omega) + \overline{\overline{\mathbf{G}}}_\mathrm{Sc}(\mathbf{r},\mathbf{r}',\omega)$, where  $\overline{\overline{\mathbf{G}}}_0(\mathbf{r},\mathbf{r}',\omega)$ represents the free-space dyadic Green's function in the absence of the dielectric bodies and $\overline{\overline{\mathbf{G}}}_\mathrm{Sc}(\mathbf{r},\mathbf{r}',\omega)$ represents the scattering dyadic Green's function originating from the presence of the dielectric bodies.

\subsection{State Vector}
To adequately include the effect of the counter-rotating interactions, we extend the Wigner-Weisskopf wave function ansatz \cite{Weisskopf1930} and use the following state vector to describe the total system \cite{Friedberg2008,Svidzinsky2010},
\begin{widetext}
\begin{align}
    \nonumber
    \ket{\Psi(t)} = & 
    \sum_{\alpha} C^{\mathrm{E_{\alpha}},\left\{0\right\}}(t) e^{-i W^{\mathrm{E_{\alpha}},\left\{0\right\}} t} \ket{\mathrm{E_{\alpha}}} \ket{\left\{ 0 \right\}} + \sum_{k=1}^{3} \int \dd{\mathrm{\bf{r}}} \int_{0}^{\infty} \dd{\omega} C^{\mathrm{G},\left\{1_k\right\}}(\mathbf{r},\omega,t) e^{-i W^{\mathrm{G},\left\{1\right\}}(\omega) t} \ket{\mathrm{G}} \ket{\left\{1_k(\mathrm{\bf{r}}, \omega)\right\}} \\
    & \quad + \sum_{\alpha} \sum_{\beta > \alpha} \sum_{k=1}^{3} \int \dd{\mathbf{r}} \int_{0}^{\infty} \dd{\omega} C^{\mathrm{E_{\alpha \beta}}, \left\{1_k\right\}}(\mathbf{r},\omega,t) e^{-i W^{\mathrm{E_{\alpha \beta}}, \left\{1\right\}}(\omega) t} \ket{\mathrm{E_{\alpha\beta}}} \ket{\left\{1_k(\mathrm{\bf{r}}, \omega) \right\}},
\label{Eq:StateVector}
\end{align}  
\end{widetext}
with
\begin{subequations}
\begin{gather}
    W^{\mathrm{E_\alpha},\{0\}} = \omega_\alpha,
    \label{Eq:EofState1} \\
    W^{\mathrm{G},\{1\}}(\omega) = \omega,
    \label{Eq:EofState2} \\
    W^{\mathrm{E_{\alpha\beta}},\{1\}}(\omega) = \omega + \omega_\alpha + \omega_\beta.
    \label{Eq:EofState3}
\end{gather}
\end{subequations}
The state vector includes the molecular (electronic) and photonic degrees of freedom in the entire system. For the molecular part, the ket state $\ket{\mathrm{G}}$ denotes that all the molecules are in their electronically ground states, i.e., $\ket{\mathrm{G}_\alpha} = \ket{\mathrm{g}_1}\ket{\mathrm{g}_2}\dots\ket{\mathrm{g}_\alpha}\dots\ket{\mathrm{g}_N}$; the ket state $\ket{\mathrm{E}_\alpha}$ denotes that $\alpha$ is in its electronically excited state while the other molecules are in their electronically ground states, i.e., $\ket{\mathrm{E}_\alpha} = \hat{\sigma}^{(+)}_\alpha \ket{\mathrm{G}}$; the ket state $\ket{\mathrm{E}_{\alpha\beta}}$ denotes that both $\alpha$ and $\beta$ are in their electronically excited states while the other molecules are in their electronically ground states, i.e., $\ket{\mathrm{E}_{\alpha\beta}} = \hat{\sigma}^{(+)}_\alpha \hat{\sigma}^{(+)}_\beta \ket{\mathrm{G}}$. For the photonic part, the ket state $\lvert\{0\}\rangle$ denotes the zero-polariton state, and the ket state $\lvert\{1_k(\mathbf{r},\omega)\}\rangle$ denotes the single-polariton state, i.e., $\lvert\{1_k(\mathbf{r},\omega)\}\rangle = \hat{f}^\dagger_k\left( \mathbf{r}, \omega \right) \lvert\{0\}\rangle$. 
$C^\mathrm{E_\alpha,\left\{0\right\}}(t)$, $C^{\mathrm{G},\{1_k\}}(\mathbf{r},\omega,t)$ and $C^\mathrm{E_{\alpha\beta},\{1_k\}}(\mathbf{r},\omega,t)$ are the probability amplitudes of $\lvert\Psi(t)\rangle$ for the states $\ket{\mathrm{E}_\alpha} \lvert\{0\}\rangle$, $\ket{\mathrm{G}} \lvert\{1_k(\mathbf{r},\omega)\}\rangle$ and $\ket{\mathrm{E}_{\alpha\beta}} \lvert\{1_k(\mathbf{r},\omega)\}\rangle
$, respectively;
$\hbar W^{\mathrm{E}_\alpha,\{0\}}$, $\hbar W^{\mathrm{G},\{1\}}(\omega)$ and $\hbar W^{\mathrm{E}_{\alpha\beta},\{1\}}(\omega)$ are the total energies of the states $\ket{\mathrm{E}_\alpha} \lvert\{0\}\rangle$, $\ket{\mathrm{G}} \lvert\{1_k(\mathbf{r},\omega)\}\rangle$ and $\ket{\mathrm{E}_{\alpha\beta}} \lvert\{1_k(\mathbf{r},\omega)\}\rangle
$, respectively. 

\subsection{Quantum Dynamics: Equation of Motion}

\begin{figure*}[t!]
    \centering
    \includegraphics[width=0.65\textwidth]{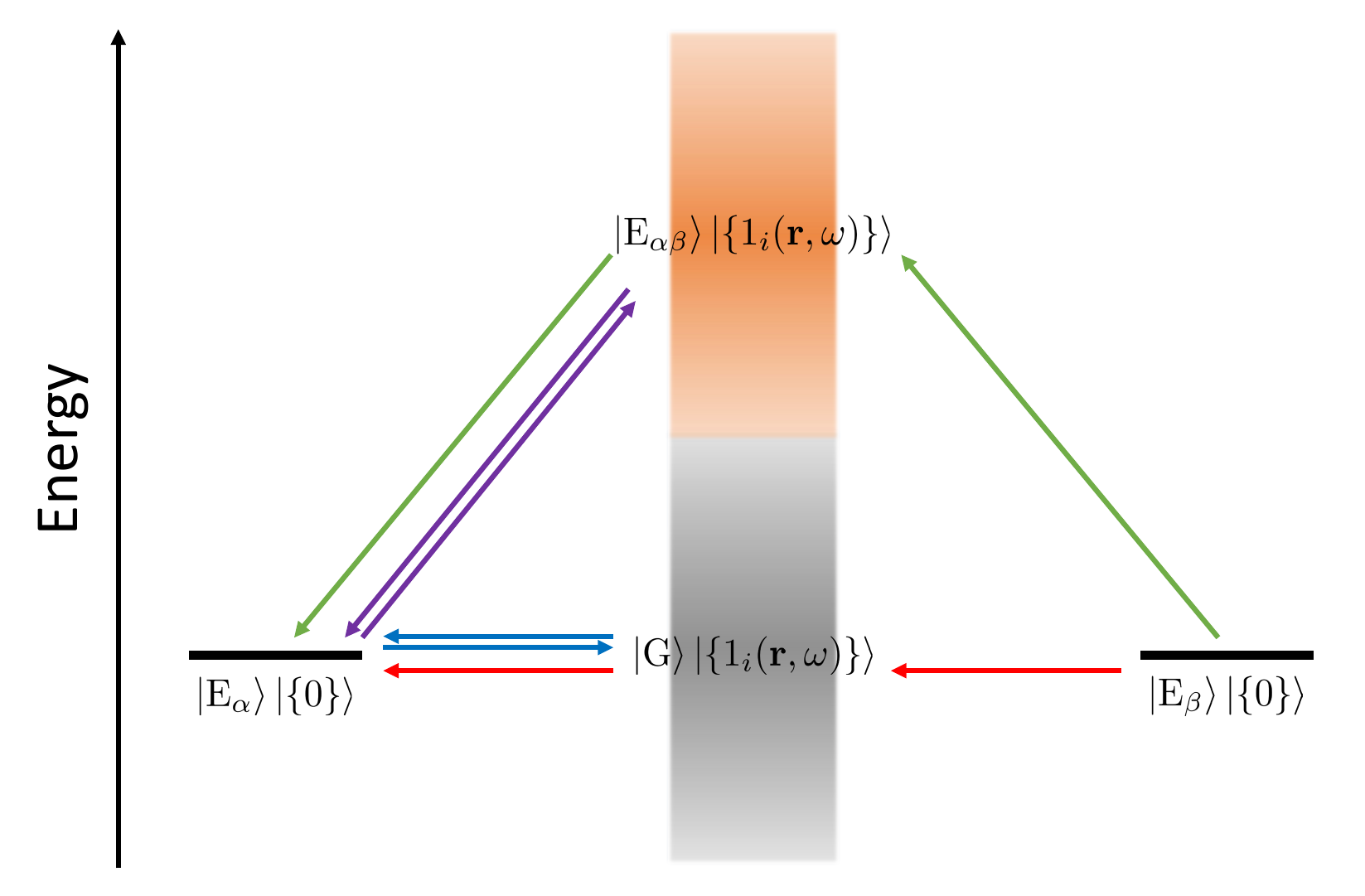}
    \caption{Schematic illustration of the quantum dynamics in Eqs.~(\ref{Eq:QuantumDynamics}).}
    \label{Fig1}
\end{figure*}

The quantum dynamics of the entire system without adopting the application of the RWA can be obtained by solving the time-dependent Schr\"{o}dinger equation $i\hbar \partial \ket{\Psi(t)}/ \partial t = \hat{H} \ket{\Psi(t)}$. After some algebra, we obtain the equation of motion as follows (see Appendix~\ref{App:DerivationOfQD} for the details),
\begin{widetext}
    \begin{align}
    \nonumber
        \dv{t} C^{\mathrm{E_{\alpha}},\left\{0\right\}}(t) = &  - \int_0^t \dd{t'} \int_0^\infty \dd{\omega} \left[ \frac{\omega^2}{\pi \hbar \varepsilon_0 c^2}\boldsymbol{\mu}^{\mathrm{eg}}_{\alpha} \cdot \mathrm{Im} \overline{\overline{\mathbf{G}}}(\mathbf{r}_\alpha,\mathbf{r}_\alpha,\omega) \cdot \boldsymbol{\mu}^{\mathrm{ge}}_{\alpha} \right] e^{-i \left( \omega - \omega_\alpha \right) t} e^{-i \left( \omega_\alpha - \omega \right) t'} C^{\mathrm{E_{\alpha}},\left\{0\right\}}(t') \\
        \nonumber
        & - \sum_{\beta \neq \alpha} \int_0^t \dd{t'} \int_0^\infty \dd{\omega} \left[ \frac{\omega^2}{\pi \hbar \varepsilon_0 c^2}\boldsymbol{\mu}^{\mathrm{eg}}_{\alpha} \cdot \mathrm{Im} \overline{\overline{\mathbf{G}}}(\mathbf{r}_\alpha,\mathbf{r}_\beta,\omega) \cdot \boldsymbol{\mu}^{\mathrm{ge}}_{\beta} \right] e^{-i \left( \omega - \omega_\alpha \right) t} e^{-i \left( \omega_\beta - \omega \right) t'} C^{\mathrm{E_{\beta}},\left\{0\right\}}(t') \\
        \nonumber
        & - \sum_{\beta\neq\alpha} \int_0^t \dd{t'} \int_0^\infty \dd{\omega} \left[ \frac{\omega^2}{\pi \hbar \varepsilon_0 c^2}\boldsymbol{\mu}^{\mathrm{ge}}_{\beta} \cdot \mathrm{Im} \overline{\overline{\mathbf{G}}}(\mathbf{r}_\beta,\mathbf{r}_\beta,\omega) \cdot \boldsymbol{\mu}^{\mathrm{eg}}_{\beta} \right] e^{-i \left( \omega + \omega_\beta \right) t} e^{-i \left( - \omega_\beta - \omega \right) t'}  C^{\mathrm{E_{\alpha}},\left\{0\right\}}(t') \\
        & - \sum_{\beta\neq\alpha} \int_0^t \dd{t'} \int_0^\infty \dd{\omega} \left[ \frac{\omega^2}{\pi \hbar \varepsilon_0 c^2}\boldsymbol{\mu}^{\mathrm{ge}}_{\beta} \cdot \mathrm{Im} \overline{\overline{\mathbf{G}}}(\mathbf{r}_\beta,\mathbf{r}_\alpha,\omega) \cdot \boldsymbol{\mu}^{\mathrm{eg}}_{\alpha} \right] e^{-i \left( \omega + \omega_\beta \right) t} e^{-i \left( - \omega_\alpha - \omega \right) t'}  C^{\mathrm{E_{\beta}},\left\{0\right\}}(t').
    \label{Eq:QuantumDynamics}
    \end{align}
\end{widetext}
The four terms in Eq.~(\ref{Eq:QuantumDynamics}) corresponds to distinct physical processes:

1. \textit{Spontaneous emission and reabsorption assisted by co-rotating interactions}: The first term on the right-hand side of Eq.~(\ref{Eq:QuantumDynamics})  corresponds to the physical process depicted by the blue arrow in FIG.~\ref{Fig1}. This process involves an initially excited molecule $\alpha$ emitting a photon and returning to its ground state. The emitted photon is then absorbed by $\alpha$ again, causing it to transition back to its excited state. In fact, this process exactly corresponds to spontaneous emission, which has been derived from previous studies \cite{Dung2000,Wang2019}.

2. \textit{Excitation energy transfer assisted by co-rotating interactions}: The second term on the right-hand side of Eq.~(\ref{Eq:QuantumDynamics}) corresponds to the physical process depicted by the red arrow in FIG.~\ref{Fig1}. In this process, an initially excited molecule $\beta$ emits a photon, transitioning back to its ground state. The emitted photon is subsequently absorbed by molecule $\alpha$, resulting in the excitation of molecule $\alpha$. This process can be regarded as excitation energy transfer assisted by co-rotating interactions \cite{Wang2022}.

3. \textit{Virtual photon emission and reabsorption assisted by counter-rotating interactions}: 
The third term on the right-hand side of Eq.~(\ref{Eq:QuantumDynamics}) corresponds to the physical process depicted by the purple arrow in FIG.~\ref{Fig1}. It involves an initially excited molecule $\alpha$ transitioning to a state where both molecules $\alpha$ and $\beta$ are excited, accompanied by the emission of a single polariton. Subsequently, the system transitions back to a state where only $\alpha$ remains excited. This process is facilitated by the counter-rotating interactions, which involve terms that describe the simultaneous creation or annihilation of both photons and molecular excitations.

\textit{4. Excitation energy transfer assisted by counter-rotating interactions}:
The fourth term on the right-hand side of Eq.~(\ref{Eq:QuantumDynamics}) corresponds to the physical process depicted by the green arrow in FIG.~\ref{Fig1}. In this process, an initially excited molecule $\beta$ transitions to a state where both molecules $\alpha$ and $\beta$ are excited, with the presence of a single polariton. Subsequently, the system transitions back to a state where only $\alpha$ remains excited. Similar to the third term, this process is facilitated by the counter-rotating interactions and can be considered as another pathway of excitation energy transfer.

\subsection{Rotating-Wave Approximation}
Under the RWA, we neglect the so-called counter-rotating terms in the interaction Hamiltonian in Eq.~(\ref{Eq:Hamiltonian_Int}). These counter-rotating terms involve processes where photons and molecular excitation are simultaneously created or annihilated. By discarding these terms, the interaction Hamiltonian is reduced to
\begin{align}
    \hat{H}^\mathrm{RWA}_\mathrm{I} &= -\sum_{\alpha=1}^{N}  \left[ \hat{\sigma}^{(+)}_{\alpha} \boldsymbol{\mu}^\mathrm{eg}_\alpha\cdot \hat{\mathbf{F}}^{(+)}(\mathbf{r}_\alpha) + \mathrm{H.c.} \right],
\label{Eq:Hamiltonian_Int_RWA}
\end{align}
and the total Hamiltonian under the RWA is defined as
\begin{align}
    \hat{H}^\mathrm{RWA} = \hat{H}_\mathrm{M} + \hat{H}_\mathrm{P} + \hat{H}^\mathrm{RWA}_\mathrm{I}.
\label{Eq:Hamiltonian_Tot_RWA}
\end{align}
Note that for the total Hamiltonian under the RWA, we do not need to consider the ket state $\ket{\mathrm{E}_{\alpha\beta}} \lvert\{1_k(\mathbf{r},\omega)\}\rangle
$ since it is not accessible without the presence of the counter-rotating interactions (assume that there is no polariton at $t=0$); therefore, we can simplify the state vector as:
\begin{widetext}
\begin{align}
    \ket{\Psi(t)}^\mathrm{RWA} = 
    \sum_{\alpha} \tilde{C}^{\mathrm{E_{\alpha}},\left\{0\right\}}(t) e^{-i W^{\mathrm{E_{\alpha}},\left\{0\right\}} t} \ket{\mathrm{E_{\alpha}}} \ket{\left\{ 0 \right\}} + \sum_{k=1}^{3} \int \dd{\mathrm{\bf{r}}} \int_{0}^{\infty} \dd{\omega} \tilde{C}^{\mathrm{G},\left\{1_k\right\}}(\mathbf{r},\omega,t) e^{-i W^{\mathrm{G},\left\{1\right\}}(\omega) t} \ket{\mathrm{G}} \ket{\left\{1_k(\mathrm{\bf{r}}, \omega)\right\}},
\label{Eq:StateVector_RWA}
\end{align}  
\end{widetext}
where $\tilde{C}^\mathrm{E_\alpha,\left\{0\right\}}(t)$ and $\tilde{C}^{\mathrm{G},\{1_k\}}(\mathbf{r},\omega,t)$ are the probability amplitudes of $\lvert\Psi(t)\rangle^\mathrm{RWA}$ for the states $\ket{\mathrm{E}_\alpha} \lvert\{0\}\rangle$ and $\ket{\mathrm{G}} \lvert\{1_k(\mathbf{r},\omega)\}\rangle$, respectively.

Solving the time-dependent Schr\"{o}dinger equation $i\hbar \partial \ket{\Psi(t)}^\mathrm{RWA}/ \partial t = \hat{H}^\mathrm{RWA} \ket{\Psi(t)}$, we obtain the equation of motion for the quantum dynamics under the RWA as follows,
\begin{widetext}
\begin{align}
    \nonumber
    \dv{t} \tilde{C}^{\mathrm{E_{\alpha}},\left\{0\right\}}(t) = & - \int_0^t \dd{t'} \int_0^\infty \dd{\omega} \left[ \frac{\omega^2}{\pi \hbar \varepsilon_0 c^2}\boldsymbol{\mu}^{\mathrm{eg}}_{\alpha} \cdot \mathrm{Im} \overline{\overline{\mathbf{G}}}(\mathbf{r}_\alpha,\mathbf{r}_\alpha,\omega) \cdot \boldsymbol{\mu}^{\mathrm{ge}}_{\alpha} \right] e^{-i \left( \omega - \omega_\alpha \right) t} e^{-i \left( \omega_\alpha - \omega \right) t'} \tilde{C}^{\mathrm{E_{\alpha}},\left\{0\right\}}(t') \\
    & - \sum_{\beta \neq \alpha} \int_0^t \dd{t'} \int_0^\infty \dd{\omega} \left[ \frac{\omega^2}{\pi \hbar \varepsilon_0 c^2}\boldsymbol{\mu}^{\mathrm{eg}}_{\alpha} \cdot \mathrm{Im} \overline{\overline{\mathbf{G}}}(\mathbf{r}_\alpha,\mathbf{r}_\beta,\omega) \cdot \boldsymbol{\mu}^{\mathrm{ge}}_{\beta} \right] e^{-i \left( \omega - \omega_\alpha \right) t} e^{-i \left( \omega_\beta - \omega \right) t'} \tilde{C}^{\mathrm{E_{\beta}},\left\{0\right\}}(t').
\label{Eq:QuantumDynamics_RWA}
\end{align}
\end{widetext}
Regarding Eq.~(\ref{Eq:QuantumDynamics_RWA}), the first and second terms on the right-hand side correspond to the physical processes represented by the blue and red arrows in FIG.~\ref{Fig1}, respectively. The physical processes depicted by the purple and green arrows are not included in Eq.~(\ref{Eq:QuantumDynamics_RWA}) due to the absence of the counter-rotating terms in the interaction Hamiltonian in Eq.~(\ref{Eq:Hamiltonian_Int_RWA}).

\section{Weak Coupling Regime: Energy Shift, Decay Rate and Dipole-Dipole Interaction}
\label{Sec:WeakCoupling}
In this section, we will show counter-rotating interactions as an essential component in describing dipole-dipole interactions, even in the weak light-matter coupling regime. To clearly demonstrate the importance of counter-rotating interactions, we further make the Markov approximation to Eq.~(\ref{Eq:QuantumDynamics}) and Eq.~(\ref{Eq:QuantumDynamics_RWA}), which correspond to the equations of motion without and with adopting the RWA, respectively, and then compare their physical meaning.

First of all, under weak light-matter interactions, we make the Markov approximation to Eq.~(\ref{Eq:QuantumDynamics}), and this equation can be simplified as (see Appendix \ref{App:DerivationOfMA}):
\begin{align}
\nonumber
    & \dv{t} C^{\mathrm{E_{\alpha}},\left\{0\right\}}(t) = \\
\nonumber
    & \quad - \frac{i}{\hbar} \left\{ \left[ \Delta_{\mathrm{e}_\alpha} + \sum_{\beta\neq\alpha} \Delta_{\mathrm{g}_\beta} \right] -i\hbar \frac{\Gamma_\alpha}{2} \right\} C^{\mathrm{E_{\alpha}},\left\{0\right\}}(t)\\
    & \quad - \frac{i}{\hbar} \sum_{\beta \neq \alpha} \mathrm{V}_\mathrm{DDI, \alpha\beta} \, e^{-i\left( \omega_\beta - \omega_\alpha \right) t} C^{\mathrm{E_{\beta}},\left\{0\right\}}(t),
\label{Eq:QuantumDynamicsMA}
\end{align}
where $\Delta_{\mathrm{e(g)}}$ represents the energy shift of the excited (ground) state of $\alpha$, $\Gamma_\alpha$ is the decay rate of the excited state of $\alpha$, and  $\mathrm{V}_\mathrm{DDI, \alpha\beta}$ denotes the effective dipole-dipole interaction between $\alpha$ and $\beta$. The energy shift $\Delta_{\mathrm{e(g)}}$ comprises two contributions: the free-space Lamb shift $\Delta^0_{\mathrm{e(g)}}$ and Casmir-Polder potential $\Delta^\mathrm{Sc}_{\mathrm{e(g)}}$, i.e.,
\begin{gather}
    \Delta_{\mathrm{e(g)}_\alpha} = \Delta^0_{\mathrm{e(g)}_\alpha} + \Delta^\mathrm{Sc}_{\mathrm{e(g)}_\alpha},
\label{Eq:EnergyShift}
\end{gather}
with
\begin{gather}
    \Delta^0_{\mathrm{e(g)}_\alpha} =
    - \mathcal{P} \int_0^\infty \dd{\omega} \frac{\omega^2}{\pi \varepsilon_0 c^2} \frac{\boldsymbol{\mu}^{\mathrm{eg}}_{\alpha} \cdot \mathrm{Im} \overline{\overline{\mathbf{G}}}_0(\mathbf{r}_\alpha,\mathbf{r}_\alpha,\omega) \cdot \boldsymbol{\mu}^{\mathrm{ge}}_{\alpha}}{\omega-(+)\omega_\alpha}, \\
    \Delta^\mathrm{Sc}_{\mathrm{e(g)}_\alpha} =
    - \mathcal{P} \int_0^\infty \dd{\omega} \frac{\omega^2}{\pi \varepsilon_0 c^2} \frac{\boldsymbol{\mu}^{\mathrm{eg}}_{\alpha} \cdot \mathrm{Im} \overline{\overline{\mathbf{G}}}_\mathrm{Sc}(\mathbf{r}_\alpha,\mathbf{r}_\alpha,\omega) \cdot \boldsymbol{\mu}^{\mathrm{ge}}_{\alpha}}{\omega-(+)\omega_\alpha},
\end{gather}
where $\mathcal{P}$ denotes the principal value. The energy shifts are identical to those derived from perturbation theory \cite{Buhmann2012,Buhmann2012b}. Note that the free-space Lamb shift $\Delta^0_{\mathrm{e(g)}_\alpha}$ is divergent in this context and requires proper treatment, such as renormalization. However, the contribution of the free-space Lamb shift is typically small after renormalization \cite{Power1959}. Therefore, for the sake of simplicity, this term can be neglected in practical calculations (or considered as being included in the transition energy of the molecule) \cite{Scheel1999,Dung2000}. The decay rate is expressed as
\begin{gather}
    \Gamma_\alpha = \frac{2 \omega_\alpha^2}{\hbar \varepsilon_0 c^2}\boldsymbol{\mu}^{\mathrm{eg}}_{\alpha} \cdot \mathrm{Im} \overline{\overline{\mathbf{G}}}(\mathbf{r}_\alpha,\mathbf{r}_\alpha,\omega_\alpha) \cdot \boldsymbol{\mu}^{\mathrm{ge}}_{\alpha},
\label{Eq:Gamma}
\end{gather}
which is consistent with the spontaneous emission rate of a molecule in a medium derived from Fermi's golden rule \cite{novotny2012principles}. The dipole-dipole interaction $\mathrm{V}_\mathrm{DDI, \alpha\beta}$ can be divided into two components, i.e.,
\begin{equation}
    \mathrm{V}_\mathrm{DDI,\alpha\beta} = \mathrm{V}_\mathrm{RDDI,\alpha\beta} + \mathrm{V}_\mathrm{ORC, \alpha\beta},
\label{Eq:V_DDI}
\end{equation}
with
\begin{equation}
    \mathrm{V}_\mathrm{RDDI,\alpha\beta} = \frac{-\omega_\beta^2}{\varepsilon_0 c^2}\boldsymbol{\mu}^{\mathrm{eg}}_{\alpha} \cdot \overline{\overline{\mathbf{G}}}(\mathbf{r}_\alpha,\mathbf{r}_\beta,\omega_\beta) \cdot \boldsymbol{\mu}^{\mathrm{ge}}_{\beta},
\label{Eq:V_RDDI}
\end{equation}

\begin{align}
    \nonumber
    \mathrm{V}_\mathrm{ORC, \alpha\beta} &= \int_0^\infty \dd{\omega} \frac{\omega^2}{\pi \varepsilon_0 c^2} \frac{\boldsymbol{\mu}^{\mathrm{eg}}_{\alpha} \cdot \mathrm{Im} \overline{\overline{\mathbf{G}}}(\mathbf{r}_\alpha,\mathbf{r}_\beta,\omega) \cdot \boldsymbol{\mu}^{\mathrm{ge}}_{\beta}}{\omega+\omega_\beta} \\
    &- \int_0^\infty \dd{\omega} \frac{\omega^2}{\pi \varepsilon_0 c^2} \frac{\boldsymbol{\mu}^{\mathrm{eg}}_{\alpha} \cdot \mathrm{Im} \overline{\overline{\mathbf{G}}}(\mathbf{r}_\alpha,\mathbf{r}_\beta,\omega) \cdot \boldsymbol{\mu}^{\mathrm{ge}}_{\beta}}{\omega+\omega_\alpha}.
\label{Eq:V_ORC}
\end{align}
$\mathrm{V}_\mathrm{RDDI,\alpha\beta}$ represents the dipole-dipole interaction between a pair of on-resonant ($\omega_\alpha = \omega_\beta$) molecules, which is identical to the resonant dipole-dipole interaction in the presence of dielectric bodies obtained using perturbation theory \cite{Dung2002}. On the other hand, $\mathrm{V}_\mathrm{ORC,\alpha\beta}$ can be regarded as a correction to the dipole-dipole interaction when the molecules are off-resonant ($\omega_\alpha \neq \omega_\beta$) since it is only non-zero when $\omega_\alpha \neq \omega_\beta$. The calculation of $\mathrm{V}_\mathrm{ORC,\alpha\beta}$ is complicated as it requires the evaluation of integrals of the dyadic Green's function over all the positive frequencies. A useful technique is to transform the integral to the imaginary axis, where the Dyadic Green's function is much better behaved \cite{Dzsotjan2011}. Using this technique, one can derive the explicit form of the free-space off-resonance correction $\mathrm{V}^0_\mathrm{ORC,\alpha\beta}$ [replace $\overline{\overline{\mathbf{G}}}(\mathbf{r}_\alpha,\mathbf{r}_\beta,\omega)$ with $\overline{\overline{\mathbf{G}}}_0(\mathbf{r}_\alpha,\mathbf{r}_\beta,\omega)$ in Eq.~(\ref{Eq:V_ORC})], as shown in Appendix~\ref{App:ImaginaryAxis}. In this work, we will not conduct a deeper discussion on the effect of the off-resonance correction; however, we would like to emphasize that this correction may play an important role when the frequency detuning between the molecular transitions is large.

Furthermore, we move on to the equation of motion with the RWA. Similarly, we make the Markov approximation to Eq.~(\ref{Eq:QuantumDynamics_RWA}), and this equation can be simplified as
\begin{align}
\nonumber
    & \dv{t} \tilde{C}^{\mathrm{E_{\alpha}},\left\{0\right\}}(t) = \\
\nonumber
    & \quad - \frac{i}{\hbar} \left[ \Delta_{\mathrm{e}_\alpha} -i\hbar \frac{\Gamma_\alpha}{2} \right] \tilde{C}^{\mathrm{E_{\alpha}},\left\{0\right\}}(t) \\
    & \quad - \frac{i}{\hbar} \sum_{\beta \neq \alpha} \tilde{\mathrm{V}}_\mathrm{DDI, \alpha\beta} \, e^{-i\left( \omega_\beta - \omega_\alpha \right) t} \tilde{C}^{\mathrm{E_{\beta}},\left\{0\right\}}(t),
\label{Eq:QuantumDynamicsMA_RWA}
\end{align}
where
\begin{equation}
    \tilde{\mathrm{V}}_\mathrm{DDI,\alpha\beta} = \mathrm{V}_\mathrm{RDDI,\alpha\beta} + \mathrm{V}_\mathrm{QC, \alpha\beta},
\label{Eq:V_DDI_RWA}
\end{equation}
\begin{equation}
    \mathrm{V}_\mathrm{QC, \alpha\beta} = \int_0^\infty \dd{\omega} \frac{\omega^2}{\pi \varepsilon_0 c^2} \frac{\boldsymbol{\mu}^{\mathrm{eg}}_{\alpha} \cdot \mathrm{Im} \overline{\overline{\mathbf{G}}}(\mathbf{r}_\alpha,\mathbf{r}_\beta,\omega) \cdot \boldsymbol{\mu}^{\mathrm{ge}}_{\beta}}{\omega+\omega_\beta}.
\label{Eq:V_QC}
\end{equation}

Comparing Eq.~(\ref{Eq:QuantumDynamicsMA}) and Eq.~(\ref{Eq:QuantumDynamicsMA_RWA}), it is obvious that the quantum dynamics without and with making the RWA under the Markov approximation exhibits two major differences. The first major difference: the energy shifts of the ground-state molecules ($\sum_{\beta\neq\alpha} \Delta_{\mathrm{g}_\beta}$) are absent in the dynamical equation under the RWA. The second major difference: from Eq.~(\ref{Eq:V_DDI}) and Eq.~(\ref{Eq:V_DDI_RWA}), one can find that the dipole-dipole interactions between a pair of molecules exhibit disparities in the two dynamical equations.

The second major difference naturally raises a question: which one is a correct form of the dipole-dipole interaction, Eq.~(\ref{Eq:V_DDI}) or Eq.~(\ref{Eq:V_DDI_RWA})? To answer this question, we conduct the following discussion. First, when the molecules are on-resonant ($\omega_\alpha = \omega_\beta$), the dipole-dipole interaction within the RWA, i.e., $\tilde{\mathrm{V}}_\mathrm{DDI,\alpha\beta}$ in Eq.~(\ref{Eq:QuantumDynamicsMA_RWA}), is expressed as a sum of the resonant dipole-dipole interaction $\mathrm{V}_\mathrm{RDDI,\alpha\beta}$ [Eq.~(\ref{Eq:V_RDDI})] and an additional correction term $\mathrm{V}_\mathrm{QC,\alpha\beta}$, which has been reported in several previous studies \cite{Dzsotjan2011,Ren2017,Varguet2021}. It is worth noting that this correction term is non-zero even when the molecules are on-resonant; therefore, $\tilde{\mathrm{V}}_\mathrm{DDI,\alpha\beta}$ always deviates from the resonant dipole-dipole interaction $\mathrm{V}_\mathrm{RDDI,\alpha\beta}$. In some previous studies, this deviation is considered an important effect that cannot be neglected in describing inter-molecule interaction \cite{Dzsotjan2011} and is regarded as a quantum correction that cannot be obtained from classical electrodynamics \cite{Varguet2021}. However, we speculate that this so-called quantum correction term may only be a product (an artifact) of the rotating-wave approximation since under the on-resonance condition, this correction no longer exists when the counter-rotating interactions are included. 
Second, when the molecules are off-resonant ($\omega_\alpha \neq \omega_\beta$), the dipole-dipole interaction without the RWA, i.e., $\mathrm{V}_\mathrm{DDI,\alpha\beta}$ in Eq.~(\ref{Eq:V_DDI}), gives a reasonable correction $\mathrm{V}_\mathrm{ORC,\alpha\beta}$ because the absolute value of $\mathrm{V}_\mathrm{ORC,\alpha\beta}$ is much smaller than that of $\mathrm{V}_\mathrm{QC,\alpha\beta}$. Third, in the short-distance (non-retarded) limit $\omega_{\beta(\alpha)} R / c \ll 1$, where $ \mathbf{r}_\alpha - \mathbf{r}_\beta \equiv R \mathbf{n}_R$, the dyadic Green's function becomes purely longitudinal \cite{Buhmann2012}, and the dipole-dipole interactions can be approximated by their free-space and electrostatic ($\omega_{\alpha(\beta)}/c \rightarrow 0$) limits; therefore,
\begin{gather}
    \mathrm{V}_\mathrm{DDI,\alpha\beta} \approx 2 \tilde{\mathrm{V}}_\mathrm{DDI,\alpha\beta} \approx \frac{\boldsymbol{\mu}^\mathrm{eg}_\alpha \cdot \boldsymbol{\mu}^\mathrm{ge}_\beta - 3 \left( \boldsymbol{\mu}^\mathrm{eg}_\alpha \cdot \mathbf{n}_R \right) \left( \boldsymbol{\mu}^\mathrm{ge}_\beta \cdot \mathbf{n}_R \right)}{4 \pi \varepsilon_0 R^3}.
\label{Eq:V_DDI_shortD}
\end{gather}
Equation~(\ref{Eq:V_DDI_shortD}) clearly shows that $\mathrm{V}_\mathrm{DDI,\alpha\beta}$ converges to the conventional Coulomb dipole-dipole interaction, while $\tilde{\mathrm{V}}_\mathrm{DDI,\alpha\beta}$ converges to only half of the Coulomb dipole-dipole interaction. The omission of half of the Coulomb dipole-dipole interaction within the framework of the RWA reinforces the significance of incorporating the counter-rotating terms in the interaction Hamiltonian. In addition to our study, in fact, the reduction of half of the dipole-dipole interaction in free space due to the use of the RWA has also been reported recently by Wubs et al \cite{Jørgensen2022}. 
Based on the above discussion, we can conclude that counter-rotating interactions play a key role even in the weak light-matter coupling regime.

In short, by examining the dynamical equation in the weak coupling regime [Eq.~(\ref{Eq:QuantumDynamicsMA})], we can obtain the energy shift, decay rate, and resonant dipole-dipole interaction of molecules in a dielectric environment, and these results are consistent with those derived from perturbation theory in previous works. This agreement not only provides robust support for the validity of our theoretical approach but also demonstrates counter-rotating interaction as an essential component for dipole-dipole interactions.

\section{Quantum Dynamics of A pair of Identical Molecules above a Plasmonic Surface}
\label{Sec:NumericalStudy}

\begin{figure}[h]
    \centering
    \includegraphics[width=0.4\textwidth]{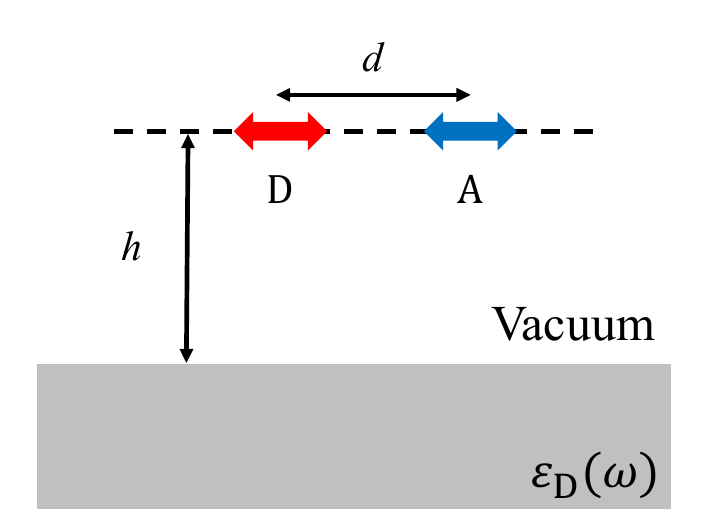}
    \caption{Schematic illustration of a donor (A) and an acceptor (A) above a plasmonic surface, where the donor-acceptor distance and the molecule-dielectric distance are given by $d$ and $h$, respectively.}
    \label{Fig2}
\end{figure}

In this section, we numerically investigate the effect of counter-rotating interactions on the quantum dynamics of multiple molecules in a complex dielectric environment. For simplicity, we adopt a minimal model consisting of a pair of identical molecules, i.e., a donor (D) and an acceptor (A), above a plasmonic surface, with their transition dipole moments parallel to the normal direction of the plasmonic surface, as depicted in FIG.~(\ref{Fig2}). We denote the distance between the donor and acceptor as $d$ and the distance between the molecules and plasmonic surface as $h$. The plasmonic surface is modeled by the following dielectric functions:
\begin{align}
    \varepsilon_\mathrm{r}(\mathbf{r},\omega) = 
    \begin{cases}
        1,    & z>0, \\
        \varepsilon_\mathrm{D}(\omega),    & z<0, 
    \end{cases} 
    \label{Eq:Dielectrics}
\end{align}
where $\varepsilon_\mathrm{D}(\omega) = 1 - 5/(\omega^2+0.1i\omega)$ is an artificial Drude model. For the molecules, the transition frequency is given by $\hbar\omega_\mathrm{D} = \hbar\omega_\mathrm{A} = 3.525$ eV, which is in resonance with the frequency of the surface plasmon polariton mode of the chosen plasmonic surface, and the magnitude of the transition dipole moment is given by $\abs{\boldsymbol{\mu}_\mathrm{D}}=\abs{\boldsymbol{\mu}_\mathrm{A}}=10$ Debye. The dyadic Green's functions of the system can be obtained through the Fresnel method \cite{chew1995}, and detailed information regarding the methodology can be found in our previous works \cite{Lee2021}.

\begin{figure*}[!t]
    \centering
    \includegraphics[width=1\textwidth]{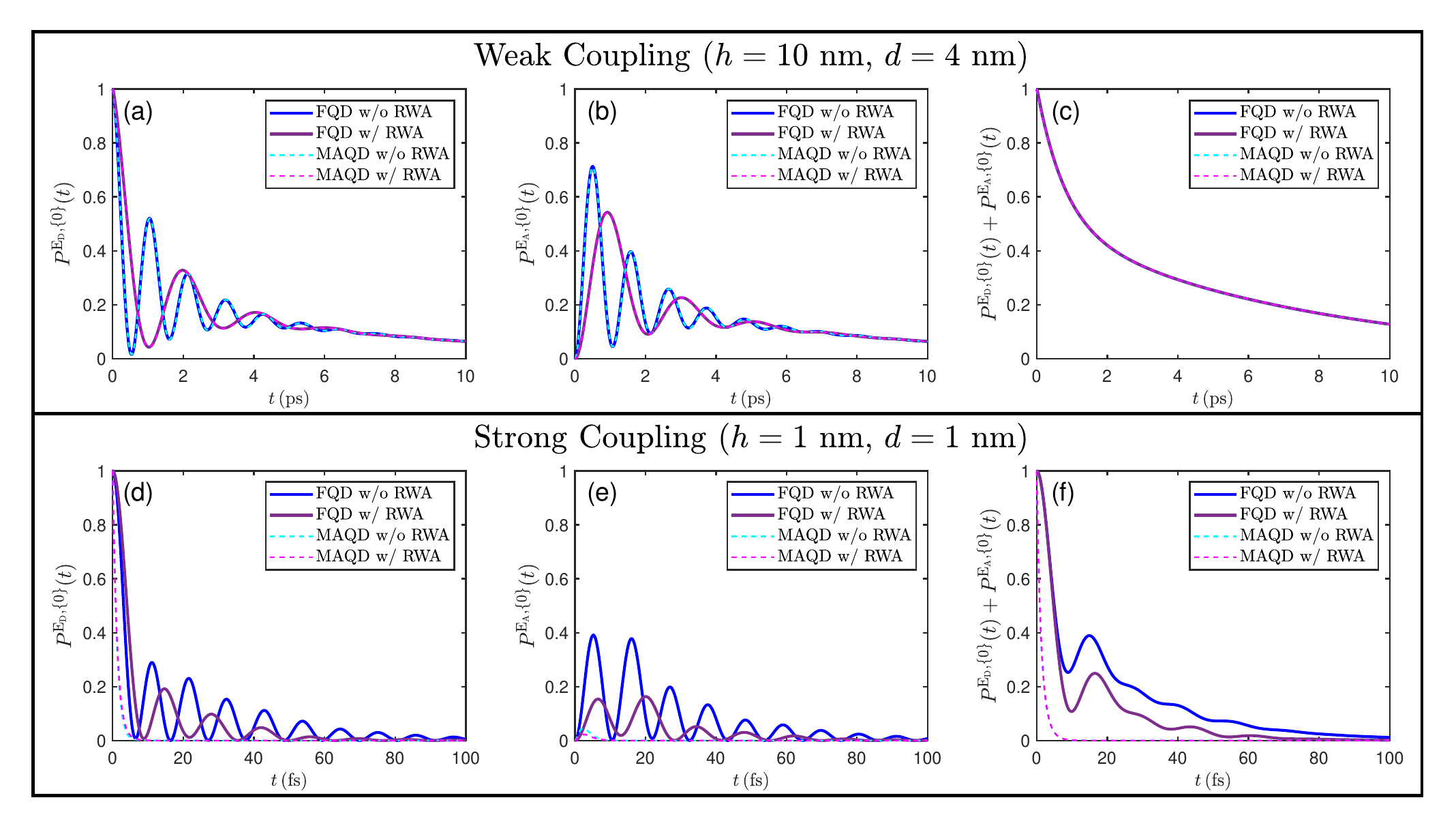}
    \caption{Excited-state population dynamics of a donor and an acceptor above a plasmonic surface. (a) Population dynamics of the donor, (b) population dynamics of the acceptor, and (c) total excited-state population dynamics when $h=10$ nm and $d=4$ nm. (d) Population dynamics of the donor, (e) population dynamics of the acceptor, and (f) total excited-state population dynamics when $h=1$ nm and $d=1$ nm. FQD and MAQD denote full quantum dynamics and Markov-approximated quantum dynamics, respectively.}
    \label{Fig3}
\end{figure*}

In order to efficiently perform numerical calculations based on Eqs.~(\ref{Eq:QuantumDynamics}), (\ref{Eq:QuantumDynamics_RWA}), (\ref{Eq:QuantumDynamicsMA}), and (\ref{Eq:QuantumDynamicsMA_RWA}), we implement additional simplifications. For Eqs.~(\ref{Eq:QuantumDynamics}) and (\ref{Eq:QuantumDynamics_RWA}), we decompose the dyadic Green's function as a sum of the free-space contribution and scattering contribution, i.e., $\overline{\overline{\mathbf{G}}}(\mathbf{r},\mathbf{r}',\omega) = \overline{\overline{\mathbf{G}}}_0(\mathbf{r},\mathbf{r}',\omega) + \overline{\overline{\mathbf{G}}}_\mathrm{Sc}(\mathbf{r},\mathbf{r}',\omega)$. Since the coupling between the molecules and the free-space field is typically weak, we can apply the Markov approximation only to the parts involving $\overline{\overline{\mathbf{G}}}_0(\mathbf{r},\mathbf{r}',\omega)$ and neglect the small free-space Lamb shift $\Delta^0_{\mathrm{e(g)}}$. This simplification leads to our working equations, as shown in Eqs.~(\ref{Eq:FQDwoRWA}) and (\ref{Eq:FQDwRWA}). These two equations will be referred to as the full quantum dynamics (FQD) without and with the RWA in the subsequent discussion. For Eqs.~(\ref{Eq:QuantumDynamicsMA}) and (\ref{Eq:QuantumDynamicsMA_RWA}), we discard the small free-space Lamb shift $\Delta^0_{\mathrm{e(g)}}$, resulting in our working equations as shown in Eqs.~(\ref{Eq:MAQDwoRWA}) and (\ref{Eq:MAQDwRWA}). These two equations will be referred to as the Markov-approximated quantum dynamics (MAQD) without and with the RWA in the subsequent discussion. The numerical study will focus on how the counter-rotating interactions affects the quantum dynamics in the weak and strong coupling conditions, where the (weak/strong) coupling condition can be identified by the consistency between the population dynamics obtained from the FQD and MAQD, e.g., the consistency of the population dynamics of FQD and MAQD indicates the weak coupling condition \cite{Wang2020b,Chuang2022}.

In FIG.~\ref{Fig3}(a)-(c), when $h=10$ nm and $d=4$ nm, the excited-state population of the donor, i.e., $P^\mathrm{E_D,\{0\}}(t)=\abs{C^\mathrm{E_D,\{0\}}(t)}^2$, the excited-state population of the acceptor, i.e., $P^\mathrm{E_A,\{0\}}(t)=\abs{C^\mathrm{E_A,\{0\}}(t)}^2$, and the total excited-state population, i.e., $P^\mathrm{E_D,\{0\}}(t) + P^\mathrm{E_A,\{0\}}(t)$, obtained via FQD and MAQD match each other; therefore, this circumstance can be identified as a weak coupling condition. In this case, the individual population dynamics of the donor $P^\mathrm{E_D,\{0\}}(t)$ and acceptor $P^\mathrm{E_A,\{0\}}(t)$ calculated without and with the application of the RWA are quite distinct from each other, as shown in FIG.~\ref{Fig3}(a) and (b). More specifically, the oscillation frequency of the population dynamics without the RWA is roughly two times the oscillation frequency of the population dynamics with the RWA. Although the individual population dynamics are different, it is surprising that the total population dynamics are still identical in this case, as shown in Fig~\ref{Fig3}(c). This phenomenon can be further confirmed by the analytical solution of MAQD (with the initial condition $C^\mathrm{E_\alpha,\{0\}}(t) = \delta_\mathrm{\alpha D}$), which can be expressed as
\begin{align}
    \begin{cases}
    P^\mathrm{E_D,\{0\}}(t) = e^{-\Gamma t} \left\{ \sinh^2 \left[ \mathrm{Im}(\mathrm{V}/\hbar)t \right] + \cos^2 \left[\mathrm{Re}(\mathrm{V}/\hbar)t \right] \right\}, \\
    P^\mathrm{E_A,\{0\}}(t) = e^{-\Gamma t} \left\{ \sinh^2 \left[\mathrm{Im}(\mathrm{V}/\hbar)t \right] + \sin^2 \left[\mathrm{Re}(\mathrm{V}/\hbar)t \right] \right\}, \\
    P^\mathrm{E_D,\{0\}}(t) + P^\mathrm{E_A,\{0\}}(t) = e^{-\Gamma t} \cosh \left[2\mathrm{Im}(\mathrm{V}/\hbar)t \right].
    \end{cases}
\label{Eq:Population_MA_ana}
\end{align}
In Eq.~(\ref{Eq:Population_MA_ana}), $\Gamma$ and $\mathrm{V}$ are defined as: (i) $\Gamma=\Gamma_\mathrm{D}=\Gamma_\mathrm{A}$ [Eq.~\ref{Eq:Gamma}] for both the population dynamics without and with the RWA. (ii) $\mathrm{V}=\mathrm{V}_\mathrm{DDI, DA} = \mathrm{V}_\mathrm{DDI, AD}$ [Eq.~(\ref{Eq:V_DDI})] for the population dynamics without the RWA, and $\mathrm{V}=\tilde{\mathrm{V}}_\mathrm{DDI, DA} = \tilde{\mathrm{V}}_\mathrm{DDI, AD}$ [Eq.~(\ref{Eq:V_DDI_RWA})] for the population dynamics with the RWA.

The analytical solution of the population dynamics in Eq.~(\ref{Eq:Population_MA_ana}) provides plenty of information. First, recall that $e^{-\Gamma t}$ is exactly the single-molecule excited-state population, and $\cosh(x)$ is greater than or equal to 1. As a result, one can conclude that the inclusion of a second molecule can slow down the decay of the total excited-state population, which is known as the subradiance. Second, the total excited-state population only depends on $\Gamma$ and $\mathrm{Im}(\mathrm{V})$, i.e., the imaginary part of the dipole-dipole interaction.  Thus, the total excited-state population is unaffected by the RWA since $\mathrm{Im}(\mathrm{V}_\mathrm{DDI, DA}) = \mathrm{Im}(\tilde{\mathrm{V}}_\mathrm{DDI, DA})$. Third, the individual excited-state population depends on both $\mathrm{Im}(\mathrm{V})$ and $\mathrm{Re}(\mathrm{V})$, where $\mathrm{Im}(\mathrm{V})$ modifies the decay behavior and $\mathrm{Re}(\mathrm{V})$ generates the oscillatory pattern. Since $\mathrm{Re}(\mathrm{V}_\mathrm{DDI, DA}) \neq \mathrm{Re}(\tilde{\mathrm{V}}_\mathrm{DDI, DA})$, the individual excited-state population is sensitive to the RWA. Moreover, the two-times oscillation frequency can be explained by the fact that $\mathrm{Re}(\mathrm{V}_\mathrm{DDI, DA}) \approx 2 \mathrm{Re}(\tilde{\mathrm{V}}_\mathrm{DDI, DA})$ in the short distance limit in Eq.~(\ref{Eq:V_DDI_shortD}).

In FIG.~\ref{Fig3}(d)-(f), when $h=1$ nm and $d=1$ nm, the population dynamics obtained via FQD and MAQD no longer match each other. In other words, we can identify this circumstance as a strong coupling condition. In this case, not only the individual population dynamics, i.e., $P^\mathrm{E_D,\{0\}}(t)$ and $P^\mathrm{E_A,\{0\}}(t)$, but also the total excited population, i.e., $P^\mathrm{E_D,\{0\}}(t)+P^\mathrm{E_A,\{0\}}(t)$, are sensitive to the RWA.

In summary, the numerical and analytical analysis of the population dynamics in the specific system reveals the following findings. (i) Under weak coupling conditions, the RWA can affect the excited-state population of individual molecules while leaving the total excited-state population unchanged (in our chosen system). However, we would like to emphasize that the total excited-state population is unaffected by the RWA in the weak coupling regime is not a general principle. In fact, in cases where the donor and acceptor are at different heights, the donor and acceptor are non-identical, or when more than two molecules are involved, the total excited-state population becomes sensitive to the use of the RWA \cite{Jørgensen2022}. (ii) Under strong coupling conditions, both the individual and total excited-state dynamics are highly sensitive to the RWA. These results further emphasize the significance of counter-rotating interactions in quantum dynamics in both strong and weak coupling regimes.

\section{Conclusion}
In this study, we investigate the influence of counter-rotating interactions on the quantum dynamics of multiple molecules in complex dielectric environments within the framework of MQED. Our general theory of quantum dynamics shows that the neglect of the counter-rotating interactions leads to missing several crucial physical processes, e.g., virtual photon emission and reabsorption. We summarize our main findings as follows. First, in the weak coupling regime, the lack of these processes leads to the absence of energy shifts in the ground-state molecules and incorrect dipole-dipole interactions between molecule pairs. Second, our study clearly demonstrates that counter-rotating interactions play an essential component in dipole-dipole interactions. Our analysis points out that within the RWA, the dipole-dipole interactions converge to only half of the conventional Coulomb interaction in the short-distance (non-retarded) limit. 
Third, our numerical simulations reveal that in the weak coupling regime, the absence of the counter-rotating interactions can significantly influence the dynamics of individual molecules while leaving the total excited-state population unchanged. Conversely, in the strong coupling regime, both individual and total excited-state dynamics exhibit sensitivity to the counter-rotating interactions. To sum up, through the analysis of the dynamical equations and a specific case study of population dynamics, we show that the counter-rotating interactions play crucial roles in both strong and weak coupling regimes. Hence, it is imperative to always exercise caution when making the rotating-wave approximation. We believe that this work will provide important insights into the study of light-matter interactions.

While we have demonstrated the significance of counter-rotating interactions, there are still unresolved issues that warrant further investigations. First, in this article, we have modeled molecules as two-level systems, neglecting the influence of other molecular excited states. However, these excited states play a crucial role in certain properties, such as energy shift calculations. Second, our analysis includes up to two molecular excitations and one polariton in the wavefunction ansatz. Nevertheless, in regimes of ultrastrong and deep-strong coupling, higher excitation states, such as three-molecule excitations and two-polariton states, may also impact quantum dynamics significantly. These issues are important for exploring the quantum dynamics of a collection of molecules coupled with quantum light. Finally, we hope that this work could inspire further investigations into emerging quantum electrodynamic phenomena in chemistry and molecular physics.

\begin{acknowledgments}
Hsu thanks Academia Sinica (AS-CDA-111-M02) and the Ministry of Science and Technology of Taiwan (110-2113-M-001-053 and 111-2113-M-001-027-MY4) for the financial support.
\end{acknowledgments}


\appendix

\section{Derivation of Eq.~(\ref{Eq:QuantumDynamics})}
\label{App:DerivationOfQD}
To derive Eq.~(\ref{Eq:QuantumDynamics}), we substitute the Hamiltonian $\hat{H}$ in Eq.~(\ref{Eq:Hamiltonian_Tot}) and the state vector in Eq.~(\ref{Eq:StateVector}) into the time-dependent Schr\"{o}dinger equation $i\hbar \partial \ket{\Psi(t)}/ \partial t = \hat{H} \ket{\Psi(t)}$, and then we obtain the following coupled differential equations,
\begin{widetext}
\begin{align}
    i\hbar \pdv{t}C^{\mathrm{G},\left\{1_k\right\}}(\mathbf{r},\omega,t) e^{-i W^{\mathrm{G},\left\{1\right\}}(\omega) t}  = - \sum_{\alpha} \left[ \boldsymbol{\mu}^{\mathrm{ge}}_{\alpha} \cdot \overline{\overline{\mathcal{G}}}\vphantom{a}^*(\mathbf{r}_\alpha,\mathbf{r},\omega) \right]_k C^{\mathrm{E_{\alpha}},\left\{0\right\}}(t) e^{-i W^{\mathrm{E}_\alpha,\left\{0 \right\}} t},
\label{Eq:Diff1}
\end{align}
\begin{align}
\nonumber
    & i\hbar \pdv{t}C^{\mathrm{E_{\alpha \beta}}, \left\{1_k\right\}}(\mathbf{r},\omega,t) e^{-i W^{\mathrm{E}_{\alpha\beta},\left\{ 1 \right\}}(\omega) t} =  \\
    &\quad -\left[ \boldsymbol{\mu}^{\mathrm{eg}}_{\beta} \cdot \overline{\overline{\mathcal{G}}}\vphantom{a}^*(\mathbf{r}_\beta,\mathbf{r},\omega) \right]_k e^{-i W^{\mathrm{E}_\alpha,\left\{0 \right\}} t} C^{\mathrm{E_{\alpha}},\left\{0\right\}}(t) - \left[ \boldsymbol{\mu}^{\mathrm{eg}}_{\alpha} \cdot \overline{\overline{\mathcal{G}}}\vphantom{a}^*(\mathbf{r}_\alpha,\mathbf{r},\omega) \right]_k e^{-i W^{\mathrm{E}_\beta,\left\{0 \right\}} t} C^{\mathrm{E_{\beta}},\left\{0\right\}}(t),
\label{Eq:Diff2}
\end{align}
\begin{align}
\nonumber
    i\hbar \dv{t} C^{\mathrm{E_{\alpha}},\left\{0\right\}}(t) e^{-i W^{\mathrm{E}_\alpha,\left\{0 \right\}} t} = & - \sum_{k=1}^{3} \int \dd{\mathbf{r}} \int_{0}^{\infty} \dd{\omega} \left[ \boldsymbol{\mu}^{\mathrm{eg}}_{\alpha} \cdot \overline{\overline{\mathcal{G}}}(\mathbf{r}_\alpha,\mathbf{r},\omega) \right]_k e^{-i W^{\mathrm{G},\left\{1\right\}}(\omega) t} C^{\mathrm{G},\left\{1_k\right\}}(\mathbf{r},\omega,t) \\ 
    & - \sum_{\beta\neq\alpha} \sum_{k=1}^{3} \int \dd{\mathbf{r}} \int_{0}^{\infty} \dd{\omega} \left[ \boldsymbol{\mu}^{\mathrm{ge}}_{\beta} \cdot \overline{\overline{\mathcal{G}}}(\mathbf{r}_\beta,\mathbf{r'},\omega) \right]_k e^{-i W^{\mathrm{E}_{\alpha\beta},\left\{ 1 \right\}}(\omega) t} C^{\mathrm{E_{\alpha \beta}}, \left\{1_k\right\}}(\mathbf{r},\omega,t),
\label{Eq:Diff3}
\end{align}
where $\left[ \mathbf{v} \right]_k$ denotes the $k$-th component of the vector $\mathbf{v}$. Note that we have ignored the two-polariton term and the three-molecule-excitation term in obtaining Eqs.~(\ref{Eq:Diff1})-(\ref{Eq:Diff3}).
Consider that there is no polariton at $t=0$, i.e., $C^{\mathrm{G},\left\{1_k\right\}}(\mathbf{r},\omega,t=0) = C^{\mathrm{E_{\alpha \beta}}, \left\{1_k\right\}}(\mathbf{r},\omega,t=0) = 0$, we can formally integrate Eqs.~(\ref{Eq:Diff1}) and (\ref{Eq:Diff2}) and obtain
\begin{align}
    C^{\mathrm{G},\left\{1_k\right\}}(\mathbf{r},\omega,t) = \frac{i}{\hbar} \sum_{\alpha} \int_0^t \dd{t'} \left[ \boldsymbol{\mu}^{\mathrm{ge}}_{\alpha} \cdot \overline{\overline{\mathcal{G}}}\vphantom{a}^*(\mathbf{r}_\alpha,\mathbf{r},\omega) \right]_k e^{-i \left( W^{\mathrm{E}_\alpha,\left\{0 \right\}} - W^{\mathrm{G},\left\{1\right\}}(\omega) \right) t'} C^{\mathrm{E_{\alpha}},\left\{0\right\}}(t'),
\label{Eq:IntDiff1}
\end{align}
\begin{align}
\nonumber
    C^{\mathrm{E_{\alpha \beta}}, \left\{1_k\right\}}(\mathbf{r},\omega,t)  = & \frac{i}{\hbar} \int_0^t \dd{t'} \left[ \boldsymbol{\mu}^{\mathrm{eg}}_{\beta} \cdot \overline{\overline{\mathcal{G}}}\vphantom{a}^*(\mathbf{r}_\beta,\mathbf{r},\omega) \right]_k e^{-i \left( W^{\mathrm{E}_\alpha,\left\{0 \right\}} - W^{\mathrm{E}_{\alpha\beta},\left\{ 1 \right\}}(\omega) \right) t'}  C^{\mathrm{E_{\alpha}},\left\{0\right\}}(t') \\
    &  + \frac{i}{\hbar} \int_0^t \dd{t'} \left[ \boldsymbol{\mu}^{\mathrm{eg}}_{\alpha} \cdot \overline{\overline{\mathcal{G}}}\vphantom{a}^*(\mathbf{r}_\alpha,\mathbf{r},\omega) \right]_k e^{-i \left( W^{\mathrm{E}_\beta,\left\{0 \right\}} - W^{\mathrm{E}_{\alpha\beta},\left\{ 1 \right\}}(\omega) \right) t'} C^{\mathrm{E_{\beta}},\left\{0\right\}}(t').
\label{Eq:IntDiff2}
\end{align}
In order to derive Eq.~(\ref{Eq:QuantumDynamics}), we substitute Eqs.~(\ref{Eq:IntDiff1}) and (\ref{Eq:IntDiff2}) into Eq.~(\ref{Eq:Diff3}), make use of the identity \cite{Buhmann2012}
\begin{align}
    \mathrm{Im}\overline{\overline{\mathbf{G}}}(\mathbf{r}, \mathbf{r'}, \omega) = \int \dd{\mathbf{s}} \frac{\omega^2 \mathrm{Im}\left[ \varepsilon_r(\mathbf{s}, \omega) \right]}{c^2} \overline{\overline{\mathbf{G}}}(\mathbf{r}, \mathbf{s}, \omega) \overline{\overline{\mathbf{G}}}\vphantom{G}^\dagger(\mathbf{r'}, \mathbf{s}, \omega),
\end{align} apply the definitions of $W^{\mathrm{E}_\alpha,\left\{0 \right\}}$, $W^{\mathrm{G},\left\{1\right\}}(\omega)$, and $W^{\mathrm{E}_{\alpha\beta},\left\{ 1 \right\}}(\omega)$ in Eqs~(\ref{Eq:EofState1})-(\ref{Eq:EofState3}), and finally obtain 
\begin{align}
\nonumber
    &\dv{t} C^{\mathrm{E_{\alpha}},\left\{0\right\}}(t) = \\
\nonumber
    & \quad - \int_0^t \dd{t'} \int_0^\infty \dd{\omega} \left[ \frac{\omega^2}{\pi \hbar \varepsilon_0 c^2}\boldsymbol{\mu}^{\mathrm{eg}}_{\alpha} \cdot \mathrm{Im} \overline{\overline{\mathbf{G}}}(\mathbf{r}_\alpha,\mathbf{r}_\alpha,\omega) \cdot \boldsymbol{\mu}^{\mathrm{ge}}_{\alpha} \right] e^{-i \left( \omega - \omega_\alpha \right) t} e^{-i \left( \omega_\alpha - \omega \right) t'} C^{\mathrm{E_{\alpha}},\left\{0\right\}}(t') \\
\nonumber
    & \quad - \sum_{\beta \neq \alpha} \int_0^t \dd{t'} \int_0^\infty \dd{\omega} \left[ \frac{\omega^2}{\pi \hbar \varepsilon_0 c^2}\boldsymbol{\mu}^{\mathrm{eg}}_{\alpha} \cdot \mathrm{Im} \overline{\overline{\mathbf{G}}}(\mathbf{r}_\alpha,\mathbf{r}_\beta,\omega) \cdot \boldsymbol{\mu}^{\mathrm{ge}}_{\beta} \right] e^{-i \left( \omega - \omega_\alpha \right) t} e^{-i \left( \omega_\beta - \omega \right) t'} C^{\mathrm{E_{\beta}},\left\{0\right\}}(t') \\
\nonumber
    & \quad - \sum_{\beta\neq\alpha} \int_0^t \dd{t'} \int_0^\infty \dd{\omega} \left[ \frac{\omega^2}{\pi \hbar \varepsilon_0 c^2}\boldsymbol{\mu}^{\mathrm{ge}}_{\beta} \cdot \mathrm{Im} \overline{\overline{\mathbf{G}}}(\mathbf{r}_\beta,\mathbf{r}_\beta,\omega) \cdot \boldsymbol{\mu}^{\mathrm{eg}}_{\beta} \right] e^{-i \left( \omega + \omega_\beta \right) t} e^{-i \left( - \omega_\beta - \omega \right) t'}  C^{\mathrm{E_{\alpha}},\left\{0\right\}}(t') \\
    & \quad - \sum_{\beta\neq\alpha} \int_0^t \dd{t'} \int_0^\infty \dd{\omega} \left[ \frac{\omega^2}{\pi \hbar \varepsilon_0 c^2}\boldsymbol{\mu}^{\mathrm{ge}}_{\beta} \cdot \mathrm{Im} \overline{\overline{\mathbf{G}}}(\mathbf{r}_\beta,\mathbf{r}_\alpha,\omega) \cdot \boldsymbol{\mu}^{\mathrm{eg}}_{\alpha} \right] e^{-i \left( \omega + \omega_\beta \right) t} e^{-i \left( - \omega_\alpha - \omega \right) t'}  C^{\mathrm{E_{\beta}},\left\{0\right\}}(t').
\end{align}
\end{widetext}

\section{Derivation of Eq.~(\ref{Eq:QuantumDynamicsMA})}
\label{App:DerivationOfMA}
In the weak coupling regime, we apply the Markov approximation to Eq.~(\ref{Eq:QuantumDynamics}), i.e., we change $C^{\mathrm{E_{\alpha}},\left\{0\right\}}(t') \rightarrow C^{\mathrm{E_{\alpha}},\left\{0\right\}}(t)$ and $\int_0^t \dd{t'} \rightarrow \int_{-\infty}^t \dd{t'}$, and make the substitution $\tau = t -t'$ ($\dd{\tau} = -\dd{t'}$); then Eq.~(\ref{Eq:QuantumDynamics}) becomes
\begin{widetext}
\begin{align}
\nonumber
    &\dv{t} C^{\mathrm{E_{\alpha}},\left\{0\right\}}(t) = \\
\nonumber
    & \quad - \left\{ \int_0^\infty \dd{\omega} \left[ \int_0^\infty \dd{\tau} e^{-i \left( \omega - \omega_\alpha \right) \tau} \right] \left[ \frac{\omega^2}{\pi \hbar \varepsilon_0 c^2}\boldsymbol{\mu}^{\mathrm{eg}}_{\alpha} \cdot \mathrm{Im} \overline{\overline{\mathbf{G}}}(\mathbf{r}_\alpha,\mathbf{r}_\alpha,\omega) \cdot \boldsymbol{\mu}^{\mathrm{ge}}_{\alpha} \right] \right\} C^{\mathrm{E_{\alpha}},\left\{0\right\}}(t) \\
\nonumber
    & \quad - \sum_{\beta \neq \alpha} \left\{ \int_0^\infty \dd{\omega} \left[ \int_0^\infty \dd{\tau} e^{-i \left( \omega - \omega_\beta \right) \tau} \right] \left[ \frac{\omega^2}{\pi \hbar \varepsilon_0 c^2}\boldsymbol{\mu}^{\mathrm{eg}}_{\alpha} \cdot \mathrm{Im} \overline{\overline{\mathbf{G}}}(\mathbf{r}_\alpha,\mathbf{r}_\beta,\omega) \cdot \boldsymbol{\mu}^{\mathrm{ge}}_{\beta} \right] \right\} e^{-i \left( \omega_\beta - \omega_\alpha \right) t} C^{\mathrm{E_{\beta}},\left\{0\right\}}(t) \\
\nonumber
    & \quad - \sum_{\beta\neq\alpha} \left\{ \int_0^\infty \dd{\omega} \left[ \int_0^\infty \dd{\tau} e^{-i \left( \omega + \omega_\beta \right) \tau} \right] \left[ \frac{\omega^2}{\pi \hbar \varepsilon_0 c^2}\boldsymbol{\mu}^{\mathrm{ge}}_{\beta} \cdot \mathrm{Im} \overline{\overline{\mathbf{G}}}(\mathbf{r}_\beta,\mathbf{r}_\beta,\omega) \cdot \boldsymbol{\mu}^{\mathrm{eg}}_{\beta} \right] \right\} C^{\mathrm{E_{\alpha}},\left\{0\right\}}(t) \\
    & \quad - \sum_{\beta\neq\alpha} \left\{ \int_0^\infty \dd{\omega} \left[ \int_0^\infty \dd{\tau} e^{-i \left( \omega + \omega_\alpha \right) \tau} \right] \left[ \frac{\omega^2}{\pi \hbar \varepsilon_0 c^2}\boldsymbol{\mu}^{\mathrm{ge}}_{\beta} \cdot \mathrm{Im} \overline{\overline{\mathbf{G}}}(\mathbf{r}_\beta,\mathbf{r}_\alpha,\omega) \cdot \boldsymbol{\mu}^{\mathrm{eg}}_{\alpha} \right] \right\} e^{-i \left( \omega_\beta - \omega_\alpha \right) t} C^{\mathrm{E_{\beta}},\left\{0\right\}}(t).
\label{Eq:DerivationOfMA1}
\end{align}
According to the Sokhotski–Plemelj theorem, we have
\begin{align}
    \int_0^\infty \dd{\tau} e^{-i \left( \omega \pm \omega_\alpha \right) \tau} = \pi \delta \left( \omega \pm \omega_\alpha \right) - i \mathcal{P} \left(\frac{1}{ \omega \pm \omega_\alpha }\right).
\label{Eq:S-Ptheorm}
\end{align}
Substituting Eq.~(\ref{Eq:S-Ptheorm}) into Eq.~(\ref{Eq:DerivationOfMA1}), we obtain
\begin{align}
\nonumber
    &\dv{t} C^{\mathrm{E_{\alpha}},\left\{0\right\}}(t) = \\
\nonumber
    & \quad - \left\{ \left[ \frac{\omega_\alpha^2}{\hbar \varepsilon_0 c^2}\boldsymbol{\mu}^{\mathrm{eg}}_{\alpha} \cdot \mathrm{Im} \overline{\overline{\mathbf{G}}}(\mathbf{r}_\alpha,\mathbf{r}_\alpha,\omega_\alpha) \cdot \boldsymbol{\mu}^{\mathrm{ge}}_{\alpha} \right] - i \mathcal{P} \int_0^\infty \dd{\omega} \left[ \frac{\omega^2}{\pi \hbar \varepsilon_0 c^2} \frac{\boldsymbol{\mu}^{\mathrm{eg}}_{\alpha} \cdot \mathrm{Im} \overline{\overline{\mathbf{G}}}(\mathbf{r}_\alpha,\mathbf{r}_\alpha,\omega) \cdot \boldsymbol{\mu}^{\mathrm{ge}}_{\alpha}}{\omega - \omega_\alpha} \right] \right\} C^{\mathrm{E_{\alpha}},\left\{0\right\}}(t) \\
\nonumber
    & \quad - \sum_{\beta \neq \alpha} \left\{ \left[ \frac{\omega_\beta^2}{\hbar \varepsilon_0 c^2}\boldsymbol{\mu}^{\mathrm{eg}}_{\alpha} \cdot \mathrm{Im} \overline{\overline{\mathbf{G}}}(\mathbf{r}_\alpha,\mathbf{r}_\beta,\omega_\beta) \cdot \boldsymbol{\mu}^{\mathrm{ge}}_{\beta} \right] - i \mathcal{P} \int_0^\infty \dd{\omega} \left[ \frac{\omega^2}{\pi \hbar \varepsilon_0 c^2} \frac{\boldsymbol{\mu}^{\mathrm{eg}}_{\alpha} \cdot \mathrm{Im} \overline{\overline{\mathbf{G}}}(\mathbf{r}_\alpha,\mathbf{r}_\beta,\omega) \cdot \boldsymbol{\mu}^{\mathrm{ge}}_{\beta}}{\omega - \omega_\beta} \right] \right\} e^{-i \left( \omega_\beta - \omega_\alpha \right) t} C^{\mathrm{E_{\beta}},\left\{0\right\}}(t) \\
\nonumber
    & \quad - \sum_{\beta\neq\alpha} \left\{ - i \mathcal{P} \int_0^\infty \dd{\omega} \left[ \frac{\omega^2}{\pi \hbar \varepsilon_0 c^2} \frac{\boldsymbol{\mu}^{\mathrm{ge}}_{\beta} \cdot \mathrm{Im} \overline{\overline{\mathbf{G}}}(\mathbf{r}_\beta,\mathbf{r}_\beta,\omega) \cdot \boldsymbol{\mu}^{\mathrm{eg}}_{\beta}}{\omega + \omega_\beta} \right] \right\} C^{\mathrm{E_{\alpha}},\left\{0\right\}}(t) \\
    & \quad - \sum_{\beta\neq\alpha} \left\{ - i \mathcal{P} \int_0^\infty \dd{\omega} \left[ \frac{\omega^2}{\pi \hbar \varepsilon_0 c^2} \frac{\boldsymbol{\mu}^{\mathrm{ge}}_{\beta} \cdot \mathrm{Im} \overline{\overline{\mathbf{G}}}(\mathbf{r}_\beta,\mathbf{r}_\alpha,\omega) \cdot \boldsymbol{\mu}^{\mathrm{eg}}_{\alpha}}{\omega + \omega_\alpha} \right] \right\} e^{-i \left( \omega_\beta - \omega_\alpha \right) t} C^{\mathrm{E_{\beta}},\left\{0\right\}}(t).
\label{Eq:DerivationOfMA2}
\end{align}
Using the following identities \cite{Buhmann2012}
\begin{align}
    \overline{\overline{\mathbf{G}}}(\mathbf{r},\mathbf{r}',-\omega^*) = \overline{\overline{\mathbf{G}}}\vphantom{G}^*(\mathbf{r},\mathbf{r}',\omega),
\label{Eq:B4}
\end{align}
\begin{align}
    \boldsymbol{\mu}_1 \cdot \overline{\overline{\mathbf{G}}}(\mathbf{r}_1,\mathbf{r}_2,\omega) \cdot \boldsymbol{\mu}_2 = \boldsymbol{\mu}_2 \cdot \overline{\overline{\mathbf{G}}}(\mathbf{r}_2,\mathbf{r}_1,\omega) \cdot \boldsymbol{\mu}_1,
\end{align}
\begin{align}
\nonumber
    \pi \omega^2 \mathrm{Re} \overline{\overline{\mathbf{G}}}(\mathbf{r},\mathbf{r}',\omega) &= \mathcal{P} \int_{-\infty}^\infty \dd{\omega} \frac{\omega^2 \mathrm{Im} \overline{\overline{\mathbf{G}}}(\mathbf{r},\mathbf{r}',\omega)}{\omega - \omega_\alpha} \\
    &= \mathcal{P} \int_0^\infty \dd{\omega} \frac{\omega^2 \mathrm{Im} \overline{\overline{\mathbf{G}}}(\mathbf{r},\mathbf{r}',\omega)}{\omega - \omega_\alpha} + \int_0^\infty \dd{\omega} \frac{\omega^2 \mathrm{Im} \overline{\overline{\mathbf{G}}}(\mathbf{r},\mathbf{r}',\omega)}{\omega + \omega_\alpha},
\end{align}
and recalling the definitions of $\Delta_{\mathrm{e(g)}_\alpha}$ in Eq~(\ref{Eq:EnergyShift}), $\Gamma_\alpha$ in Eq~(\ref{Eq:Gamma}), and $\mathrm{V}_\mathrm{DDI,\alpha\beta}$ in Eq~(\ref{Eq:V_DDI}), we can transform Eq~(\ref{Eq:DerivationOfMA2}) into
\begin{gather}
    \dv{t} C^{\mathrm{E_{\alpha}},\left\{0\right\}}(t) = - \frac{i}{\hbar} \left\{ \left[ \Delta_{\mathrm{e}_\alpha} + \sum_{\beta\neq\alpha} \Delta_{\mathrm{g}_\beta} \right] -i\hbar \frac{\Gamma_\alpha}{2} \right\} C^{\mathrm{E_{\alpha}},\left\{0\right\}}(t) - \frac{i}{\hbar} \sum_{\beta \neq \alpha} \mathrm{V}_\mathrm{DDI, \alpha\beta} \, e^{-i\left( \omega_\beta - \omega_\alpha \right) t} C^{\mathrm{E_{\beta}},\left\{0\right\}}(t).
\end{gather}
\end{widetext}

\section{Calculation of $\mathrm{V}_\mathrm{ORC, \alpha\beta}$ and $\mathrm{V}_\mathrm{QC, \alpha\beta}$ on the Imaginary Axis}
\label{App:ImaginaryAxis}
\begin{figure}[!t]
    \centering
    \includegraphics[width=0.4\textwidth]{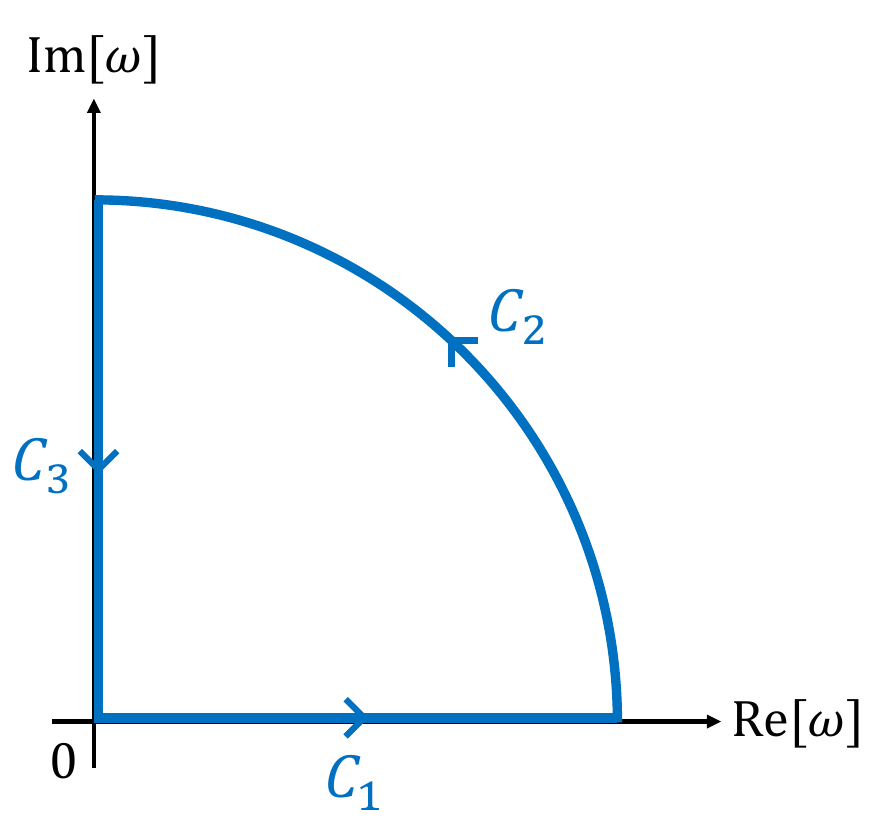}
    \caption{The contour adopted in the integral. The total contour $C$ is equal to $C_1+C_2+C_3$. Note that there is no singularity inside the contour.}
    \label{Fig4}
\end{figure}

To obtain $\mathrm{V}_\mathrm{ORC, \alpha\beta}$ and $\mathrm{V}_\mathrm{QC, \alpha\beta}$, we need to evaluate the integral

\begin{align}
    I = \int_0^\infty \dd{\omega} \frac{\omega^2}{\pi \varepsilon_0 c^2} \frac{\boldsymbol{\mu}^{\mathrm{eg}}_{\alpha} \cdot \mathrm{Im} \overline{\overline{\mathbf{G}}}(\mathbf{r}_\alpha,\mathbf{r}_\beta,\omega) \cdot \boldsymbol{\mu}^{\mathrm{ge}}_{\beta}}{\omega+\omega'},
\label{Integral}
\end{align}
where $\omega' = \omega_\alpha$ or $\omega_\beta$. $I$ can be alternatively expressed as:
\begin{align}
    I = \frac{1}{\pi \varepsilon_0} \boldsymbol{\mu}^{\mathrm{eg}}_{\alpha} \cdot  \mathrm{Im} \left[ \int_0^\infty \dd{\omega} \frac{\omega^2}{c^2} \frac{\overline{\overline{\mathbf{G}}}(\mathbf{r}_\alpha,\mathbf{r}_\beta,\omega)}{\omega+\omega'} \right] \cdot \boldsymbol{\mu}^{\mathrm{ge}}_{\beta}. 
\end{align}
For simplicity, we define $f(\omega) = \frac{\omega^2}{c^2} \frac{\overline{\overline{\mathbf{G}}}(\mathbf{r}_\alpha,\mathbf{r}_\beta,\omega)}{\omega+\omega'}$. The integral in the square bracket can be evaluated using the contour integral technique, i.e.,
\begin{align}
    \int_0^\infty \dd{\omega} f(\omega) = \oint_C \dd{\omega} f(\omega) - \int_{C_2} \dd{\omega} f(\omega) - \int_{C_3} \dd{\omega} f(\omega),
\label{Eq:Contour}
\end{align}
where the contour is shown in FIG.~\ref{Fig4}. The first term on the right-hand side of Eq.~(\ref{Eq:Contour}) is zero since there is no singularity inside the contour $C$. The second term on the right-hand side of Eq.~(\ref{Eq:Contour}) is also zero due to the asymptotic behavior of the two-point dyadic Green’s function \cite{Buhmann2012}
\begin{align}
    \lim_{\abs{\omega} \rightarrow \infty} \left.\frac{\omega^2}{c^2} \overline{\overline{\mathbf{G}}}(\mathbf{r}_\alpha,\mathbf{r}_\beta,\omega)\right|_{\mathbf{r}_\alpha \neq \mathbf{r}_\beta} = \overline{\overline{\mathbf{0}}}.
\end{align}

Therefore, we have
\begin{align}
\nonumber
    \int_0^\infty \dd{\omega} f(\omega) &= -\int_{i\infty}^0 \dd{\omega} f(\omega) \\
\nonumber
    & = - \int_{i\infty}^0 \dd{\omega} \frac{\omega^2}{c^2} \frac{\overline{\overline{\mathbf{G}}}(\mathbf{r}_\alpha,\mathbf{r}_\beta,\omega)}{\omega+\omega'} \\
\nonumber
    & = - \int_{i\infty}^0 \dd{\omega} \frac{\omega^2 \left(\omega-\omega'\right)}{c^2} \frac{\overline{\overline{\mathbf{G}}}(\mathbf{r}_\alpha,\mathbf{r}_\beta,\omega)}{\omega^2-\omega'^2} \\
\nonumber
    & = - \int_{i\infty}^0 \dd{\omega} \frac{\omega^3}{c^2} \frac{\overline{\overline{\mathbf{G}}}(\mathbf{r}_\alpha,\mathbf{r}_\beta,\omega)}{\omega^2-\omega'^2} \\
    & \quad + \int_{i\infty}^0 \dd{\omega} \frac{\omega'\omega^2}{c^2} \frac{\overline{\overline{\mathbf{G}}}(\mathbf{r}_\alpha,\mathbf{r}_\beta,\omega)}{\omega^2-\omega'^2}.
\end{align}
Making the substitution $\kappa = -i \omega$,
\begin{align}
\nonumber
    \int_0^\infty \dd{\omega} f(\omega) &= - \int_0^\infty \dd{\kappa} \frac{\kappa^3}{c^2} \frac{\overline{\overline{\mathbf{G}}}(\mathbf{r}_\alpha,\mathbf{r}_\beta,i\kappa)}{\kappa^2+\omega'^2} \\
    & \quad - i \int_0^\infty \dd{\kappa} \frac{\omega'\kappa^2}{c^2} \frac{\overline{\overline{\mathbf{G}}}(\mathbf{r}_\alpha,\mathbf{r}_\beta,i\kappa)}{\kappa^2+\omega'^2}.
\end{align}
Using Eq.~(\ref{Eq:B4}), we can obtain that the dyadic Green's function is purely real on the imaginary axis, i.e., $\overline{\overline{\mathbf{G}}}(\mathbf{r}_\alpha,\mathbf{r}_\beta,i\kappa) = \mathrm{Re}\overline{\overline{\mathbf{G}}}(\mathbf{r}_\alpha,\mathbf{r}_\beta,i\kappa)$, and arrive at
\begin{align}
    \mathrm{Im} \left[ \int_0^\infty \dd{\omega} f(\omega) \right] =  - \int_0^\infty \dd{\kappa} \frac{\omega'\kappa^2}{c^2} \frac{\mathrm{Re}\overline{\overline{\mathbf{G}}}(\mathbf{r}_\alpha,\mathbf{r}_\beta,i\kappa)}{\kappa^2+\omega'^2}.
\end{align}
Now we can express the integral $I$ as \cite{Dzsotjan2011,Varguet2021}:
\begin{align}
    I = - \int_0^\infty \dd{\kappa} \frac{\omega'\kappa^2}{\pi \varepsilon_0c^2} \frac{\boldsymbol{\mu}^{\mathrm{eg}}_{\alpha} \cdot \mathrm{Re}\overline{\overline{\mathbf{G}}}(\mathbf{r}_\alpha,\mathbf{r}_\beta,i\kappa) \cdot \boldsymbol{\mu}^{\mathrm{ge}}_{\beta} }{\kappa^2+\omega'^2}.
\label{Eq:ImaginaryAxis}
\end{align}
Eq.~(\ref{Eq:ImaginaryAxis}) is a powerful tool in cases where the dyadic Green's function on the imaginary axis is available since $\overline{\overline{\mathbf{G}}}(\mathbf{r}_\alpha,\mathbf{r}_\beta,i\kappa)$ decays rapidly \cite{Dzsotjan2011}. However, in complex dielectric environments, dyadic Green's function on the imaginary axis is difficult to obtain, and the evaluation of $I$ through Eq.~(\ref{Integral}) is more convenient.

In free space, we have the explicit expression of the two-point dyadic Green's function, which reads:
\begin{align}
\nonumber
    & \left. \overline{\overline{\mathbf{G}}}_{{\mathrm{0}}}(\mathbf{r}_\alpha,\mathbf{r}_\beta,\omega) \right|_{\mathbf{r}_\alpha \neq \mathbf{r}_\beta} = \\
\nonumber
    & \quad \frac{e^{ik_0R}}{4\pi R} \Biggl\{ \vphantom{\frac{e^R}{R}} \left(\overline{\overline{\mathbf{I}}}_3 - \mathbf{n}_R \otimes \mathbf{n}_R \right) \\
    & \quad \quad + \left(3 \mathbf{n}_R \otimes \mathbf{n}_R - \overline{\overline{\mathbf{I}}}_3\right)\left[\frac{1}{(k_0R)^{2}}-\frac{i}{k_0R} \right] \Biggl\},
\label{Eq:G0}
\end{align}
where $k_0 = \omega / c$ and $ \mathbf{r}_\alpha - \mathbf{r}_\beta \equiv R \mathbf{n}_R$. Therefore, we can evaluate the integral using Eq.~(\ref{Eq:ImaginaryAxis}). Inserting Eq.~(\ref{Eq:G0}) into Eq.~(\ref{Eq:ImaginaryAxis}) and making the substitution $x = \kappa R/c$, we obtain
\begin{align}
\nonumber
    I^0 &\equiv - \int_0^\infty \dd{\kappa} \frac{\omega'\kappa^2}{\pi \varepsilon_0c^2} \frac{\boldsymbol{\mu}^{\mathrm{eg}}_{\alpha} \cdot \mathrm{Re}\overline{\overline{\mathbf{G}}}_0(\mathbf{r}_\alpha,\mathbf{r}_\beta,i\kappa) \cdot \boldsymbol{\mu}^{\mathrm{ge}}_{\beta}}{\kappa^2+\omega'^2} \\
\nonumber
    & = - \frac{\omega'}{4 \pi^2 \varepsilon_0 c R^2} \Bigl\{ \left[ \boldsymbol{\mu}^\mathrm{eg}_\alpha \cdot \boldsymbol{\mu}^\mathrm{ge}_\beta - \left( \boldsymbol{\mu}^\mathrm{eg}_\alpha \cdot \mathbf{n}_R \right) \left( \boldsymbol{\mu}^\mathrm{ge}_\beta \cdot \mathbf{n}_R \right) \right] \mathcal{I}_1 \\
    & \quad + \left[ \boldsymbol{\mu}^\mathrm{eg}_\alpha \cdot \boldsymbol{\mu}^\mathrm{ge}_\beta - 3 \left( \boldsymbol{\mu}^\mathrm{eg}_\alpha \cdot \mathbf{n}_R \right) \left( \boldsymbol{\mu}^\mathrm{ge}_\beta \cdot \mathbf{n}_R \right) \right] \left( \mathcal{I}_2 + \mathcal{I}_3 \right) \Bigl\}.
\end{align}
$\mathcal{I}_1$, $\mathcal{I}_2$, and $\mathcal{I}_3$ are the auxiliary integrals defined as
\begin{subequations}
\begin{align}
    \mathcal{I}_1 = \int_0^\infty \dd{x} \frac{x^2 e^{-x}}{x^2 + x'^2}, \\
    \mathcal{I}_2 = \int_0^\infty \dd{x} \frac{x e^{-x}}{x^2 + x'^2}, \\
    \mathcal{I}_3 = \int_0^\infty \dd{x} \frac{e^{-x}}{x^2 + x'^2},
\end{align}
\end{subequations}
where $x' = \omega'R/c$. These auxiliary integrals can be explicitly expressed in terms of the trigonometric functions and trigonometric integrals as \cite{Gradshteyn2015}
\begin{subequations}
\begin{align}
    \mathcal{I}_1 &= -x' \left[ \mathrm{ci}(x') \sin(x') - \mathrm{si}(x') \cos(x') \right] + 1, \\
    \mathcal{I}_2 &= - \mathrm{ci}(x') \cos(x') - \mathrm{si}(x') \sin(x'), \\
    \mathcal{I}_3 &= \frac{1}{x'} \left[ \mathrm{ci}(x') \sin(x') - \mathrm{si}(x') \cos(x') \right],
\end{align}
\end{subequations}
where the trigonometric integrals follow the definitions:
\begin{align}
\nonumber
    \mathrm{ci}(x') = - \int_{x'}^{\infty} \dd{x} \frac{\cos(x)}{x}, \,\, \mathrm{si}(x') = - \int_{x'}^{\infty} \dd{x} \frac{\sin(x)}{x}.
\end{align}
The explicit form of $I^0$ is useful for the calculation of quantum dynamics. In addition, the asymptotic behavior of $I^0$, $V^0_{\mathrm{ORC},\alpha\beta}$, and $V^0_{\mathrm{QC},\alpha\beta}$ can be analyzed through the expansion of the trigonometric integrals.

\section{Numerical Implementation of FQD and MAQD}
To numerically implement FQD without the RWA, we start from Eq.~(\ref{Eq:QuantumDynamics}), separate $\overline{\overline{\mathbf{G}}}(\mathbf{r},\mathbf{r}',\omega)$ as $\overline{\overline{\mathbf{G}}}_0(\mathbf{r},\mathbf{r}',\omega) + \overline{\overline{\mathbf{G}}}_\mathrm{Sc}(\mathbf{r},\mathbf{r}',\omega)$, apply the Markov approximation only to the part involve $\overline{\overline{\mathbf{G}}}_0(\mathbf{r},\mathbf{r}',\omega)$, and discard the free-space Lamb shift $\Delta^0_\mathrm{e(g)_\alpha}$. As a result, one can numerically calculate FQD without the RWA instead of using Eq.~(\ref{Eq:QuantumDynamics}) as follows,
\begin{widetext}
\begin{align}
\nonumber
    \dv{t} C^{\mathrm{E_{\alpha}},\left\{0\right\}}(t) = & - \frac{\Gamma^0_\alpha}{2} C^{\mathrm{E_{\alpha}},\left\{0\right\}}(t) - \frac{i}{\hbar} \sum_{\beta \neq \alpha} \mathrm{V}^0_\mathrm{DDI, \alpha\beta} \, e^{-i\left( \omega_\beta - \omega_\alpha \right) t} C^{\mathrm{E_{\beta}},\left\{0\right\}}(t) \\
\nonumber
    &  - \int_0^t \dd{t'} \int_0^\infty \dd{\omega} \left[ \frac{\omega^2}{\pi \hbar \varepsilon_0 c^2}\boldsymbol{\mu}^{\mathrm{eg}}_{\alpha} \cdot \mathrm{Im} \overline{\overline{\mathbf{G}}}_\mathrm{Sc}(\mathbf{r}_\alpha,\mathbf{r}_\alpha,\omega) \cdot \boldsymbol{\mu}^{\mathrm{ge}}_{\alpha} \right] e^{-i \left( \omega - \omega_\alpha \right) t} e^{-i \left( \omega_\alpha - \omega \right) t'} C^{\mathrm{E_{\alpha}},\left\{0\right\}}(t') \\
\nonumber
    & - \sum_{\beta \neq \alpha} \int_0^t \dd{t'} \int_0^\infty \dd{\omega} \left[ \frac{\omega^2}{\pi \hbar \varepsilon_0 c^2}\boldsymbol{\mu}^{\mathrm{eg}}_{\alpha} \cdot \mathrm{Im} \overline{\overline{\mathbf{G}}}_\mathrm{Sc}(\mathbf{r}_\alpha,\mathbf{r}_\beta,\omega) \cdot \boldsymbol{\mu}^{\mathrm{ge}}_{\beta} \right] e^{-i \left( \omega - \omega_\alpha \right) t} e^{-i \left( \omega_\beta - \omega \right) t'} C^{\mathrm{E_{\beta}},\left\{0\right\}}(t') \\
\nonumber
    & - \sum_{\beta\neq\alpha} \int_0^t \dd{t'} \int_0^\infty \dd{\omega} \left[ \frac{\omega^2}{\pi \hbar \varepsilon_0 c^2}\boldsymbol{\mu}^{\mathrm{ge}}_{\beta} \cdot \mathrm{Im} \overline{\overline{\mathbf{G}}}_\mathrm{Sc}(\mathbf{r}_\beta,\mathbf{r}_\beta,\omega) \cdot \boldsymbol{\mu}^{\mathrm{eg}}_{\beta} \right] e^{-i \left( \omega + \omega_\beta \right) t} e^{-i \left( - \omega_\beta - \omega \right) t'}  C^{\mathrm{E_{\alpha}},\left\{0\right\}}(t') \\
    & - \sum_{\beta\neq\alpha} \int_0^t \dd{t'} \int_0^\infty \dd{\omega} \left[ \frac{\omega^2}{\pi \hbar \varepsilon_0 c^2}\boldsymbol{\mu}^{\mathrm{ge}}_{\beta} \cdot \mathrm{Im} \overline{\overline{\mathbf{G}}}_\mathrm{Sc}(\mathbf{r}_\beta,\mathbf{r}_\alpha,\omega) \cdot \boldsymbol{\mu}^{\mathrm{eg}}_{\alpha} \right] e^{-i \left( \omega + \omega_\beta \right) t} e^{-i \left( - \omega_\alpha - \omega \right) t'}  C^{\mathrm{E_{\beta}},\left\{0\right\}}(t'), 
\label{Eq:FQDwoRWA}
\end{align}
where $\Gamma^0_\alpha = \frac{2 \omega_\alpha^2}{\hbar \varepsilon_0 c^2}\boldsymbol{\mu}^{\mathrm{eg}}_{\alpha} \cdot \mathrm{Im} \overline{\overline{\mathbf{G}}}_0(\mathbf{r}_\alpha,\mathbf{r}_\alpha,\omega_\alpha) \cdot \boldsymbol{\mu}^{\mathrm{ge}}_{\alpha} = \frac{ \abs{\boldsymbol{\mu}^{\mathrm{ge}}_{\alpha}}^2 \omega_\alpha^3}{3 \pi \hbar \varepsilon_0 c^3}$ is the spontaneous emission rate in free space, and $\mathrm{V}^0_\mathrm{DDI, \alpha\beta}$ is obtained by substituting $\overline{\overline{\mathbf{G}}}(\mathbf{r}_\alpha,\mathbf{r}_\beta,\omega)$ with $\overline{\overline{\mathbf{G}}}_0(\mathbf{r}_\alpha,\mathbf{r}_\beta,\omega)$ in the expression of $\mathrm{V}_\mathrm{DDI, \alpha\beta}$.

To numerically implement FQD with the RWA, we start from Eq.~(\ref{Eq:QuantumDynamics_RWA}) and adopt the same procedure as in the above. As a result, we can numerically calculate FQD with the RWA instead of Eq.~(\ref{Eq:QuantumDynamics_RWA}) as follows,
\begin{align}
\nonumber
    \dv{t} \tilde{C}^{\mathrm{E_{\alpha}},\left\{0\right\}}(t) = & - \frac{\Gamma^0_\alpha}{2} \tilde{C}^{\mathrm{E_{\alpha}},\left\{0\right\}}(t) - \frac{i}{\hbar} \sum_{\beta \neq \alpha} \tilde{\mathrm{V}}^0_\mathrm{DDI, \alpha\beta} \, e^{-i\left( \omega_\beta - \omega_\alpha \right) t} \tilde{C}^{\mathrm{E_{\beta}},\left\{0\right\}}(t) \\
\nonumber
    &  - \int_0^t \dd{t'} \int_0^\infty \dd{\omega} \left[ \frac{\omega^2}{\pi \hbar \varepsilon_0 c^2}\boldsymbol{\mu}^{\mathrm{eg}}_{\alpha} \cdot \mathrm{Im} \overline{\overline{\mathbf{G}}}_\mathrm{Sc}(\mathbf{r}_\alpha,\mathbf{r}_\alpha,\omega) \cdot \boldsymbol{\mu}^{\mathrm{ge}}_{\alpha} \right] e^{-i \left( \omega - \omega_\alpha \right) t} e^{-i \left( \omega_\alpha - \omega \right) t'} \tilde{C}^{\mathrm{E_{\alpha}},\left\{0\right\}}(t') \\
    & - \sum_{\beta \neq \alpha} \int_0^t \dd{t'} \int_0^\infty \dd{\omega} \left[ \frac{\omega^2}{\pi \hbar \varepsilon_0 c^2}\boldsymbol{\mu}^{\mathrm{eg}}_{\alpha} \cdot \mathrm{Im} \overline{\overline{\mathbf{G}}}_\mathrm{Sc}(\mathbf{r}_\alpha,\mathbf{r}_\beta,\omega) \cdot \boldsymbol{\mu}^{\mathrm{ge}}_{\beta} \right] e^{-i \left( \omega - \omega_\alpha \right) t} e^{-i \left( \omega_\beta - \omega \right) t'} \tilde{C}^{\mathrm{E_{\beta}},\left\{0\right\}}(t'),
\label{Eq:FQDwRWA}
\end{align}
where $\tilde{\mathrm{V}}^0_\mathrm{DDI, \alpha\beta}$ is obtained by substituting $\overline{\overline{\mathbf{G}}}(\mathbf{r}_\alpha,\mathbf{r}_\beta,\omega)$ with $\overline{\overline{\mathbf{G}}}_0(\mathbf{r}_\alpha,\mathbf{r}_\beta,\omega)$ in the expression of $\tilde{\mathrm{V}}_\mathrm{DDI, \alpha\beta}$.

To numerically implement MAQD without the RWA, we discard the free-space Lamb shift $\Delta^0_\mathrm{e(g)_\alpha}$ in Eq.~(\ref{Eq:QuantumDynamicsMA}) and finally obtain
\begin{gather}
    \dv{t} C^{\mathrm{E_{\alpha}},\left\{0\right\}}(t) = - \frac{i}{\hbar} \left\{ \left[ \Delta^\mathrm{Sc}_{\mathrm{e}_\alpha} + \sum_{\beta\neq\alpha} \Delta^\mathrm{Sc}_{\mathrm{g}_\beta} \right] -i\hbar \frac{\Gamma_\alpha}{2} \right\} C^{\mathrm{E_{\alpha}},\left\{0\right\}}(t) - \frac{i}{\hbar} \sum_{\beta \neq \alpha} \mathrm{V}_\mathrm{DDI, \alpha\beta} \, e^{-i\left( \omega_\beta - \omega_\alpha \right) t} C^{\mathrm{E_{\beta}},\left\{0\right\}}(t).
\label{Eq:MAQDwoRWA}
\end{gather}

To numerically implement MAQD with the RWA, we discard the free-space Lamb shift $\Delta^0_\mathrm{e_\alpha}$ in Eq.~(\ref{Eq:QuantumDynamicsMA_RWA}) and finally arrive at
\begin{gather}
 \dv{t} \tilde{C}^{\mathrm{E_{\alpha}},\left\{0\right\}}(t) = - \frac{i}{\hbar} \left[ \Delta^\mathrm{Sc}_{\mathrm{e}_\alpha} -i\hbar \frac{\Gamma_\alpha}{2} \right] \tilde{C}^{\mathrm{E_{\alpha}},\left\{0\right\}}(t) - \frac{i}{\hbar} \sum_{\beta \neq \alpha} \tilde{\mathrm{V}}_\mathrm{DDI, \alpha\beta} \, e^{-i\left( \omega_\beta - \omega_\alpha \right) t} \tilde{C}^{\mathrm{E_{\beta}},\left\{0\right\}}(t).
\label{Eq:MAQDwRWA}
\end{gather}

\end{widetext}


%

\end{document}